\documentclass[usenatbib]{mn2e}
\usepackage{graphicx}
\usepackage{soul}

\newcommand{\Bran}{B_{ran}}

\newcommand{\microG}{$\mu G$\ }

\newcommand{\Bxp}{B_x^\prime}

\newcommand{\Bzp}{B_z^\prime}

\newcommand{\mnras} {MNRAS}
\newcommand{\apjl}{ApJL}
\newcommand{\apj}{ApJ}
\newcommand{\ssr}{Space Sci. Rev.}
\newcommand{\apjs}{ApJS}
\newcommand{\prd}{PhRvD}
\newcommand{\aap}{A\&A}
\newcommand{\nat}{Nature}
\newcommand{\jcap}{JCAP}
\newcommand{\aaps}{A\&AS}

\title[Interstellar synchrotron emission]{Galactic synchrotron emission with cosmic ray propagation models}

\author[Elena Orlando and Andrew W. Strong] {Elena Orlando$^{1}$\thanks{E-mail: 
eorlando@stanford.edu; aws@mpe.mpg.de}, Andrew Strong$^{2}$\footnotemark[1]\\
$^{1}$W.W. Hansen Experimental Physics Laboratory, Kavli Institute for Particle
Astrophysics and Cosmology,\\ 
Stanford University, Stanford, CA 94305, USA\\
$^{2}$Max-Planck-Institut f\"ur extraterrestrische Physik,
Postfach 1312, D-85741 Garching, Germany}

\begin{document}

\date{Accepted 2013 September 10. Received 2013 September 5; in original form 2013 July 29}

\pagerange{\pageref{firstpage}--\pageref{lastpage}} \pubyear{2002}

\maketitle
\label{firstpage}

\begin{abstract}
Cosmic ray (CR) leptons  produce radio synchrotron radiation by gyrating in interstellar magnetic fields (B-field). 
Details of B-fields, CR electron distributions and  propagation are still uncertain.\\
We present  developments in our modelling of Galactic radio emission with the GALPROP code.
It now includes calculations of radio polarization, absorption and free-free emission. 
Total and polarized synchrotron emission are investigated in the context of physical models of CR propagation.
Predictions are compared with radio data from 22 MHz to 2.3 GHz, and {\it Wilkinson Microwave Anisotropy Probe} data at 23 GHz. 
Spatial and spectral effects on the synchrotron modelling with different CR distribution, propagation halo size and CR propagation models are presented.
We find that all-sky total intensity and polarization maps are reasonably reproduced by including an anisotropic B-field, with comparable intensity to the regular one defined by rotation measures.   
A halo size of 10 kpc, which is larger than usually assumed, is favoured.\\
This work provides a basis for further studies on foreground emission with the {\it Planck} satellite and on interstellar gamma-ray emission with {\it Fermi}-Large Area Telescope.
\end{abstract}


\section{Introduction} \label{sec:introduction}

Galactic synchrotron radiation is produced by cosmic ray (CR) electrons and positrons propagating in  interstellar magnetic fields (B-fields). 
Synchrotron emission is one of the major Galactic components from several hundred MHz to several hundred GHz. 
In the microwave band free-free and dust emission tend to dominate making more difficult the separation of the components. 
Advanced modelling of the different emissions, both total and polarized components is important for separating synchrotron emission from other components. 
Synchrotron modelling requires a knowledge of the Galactic B-fields and CR electrons in the Galaxy. 
Hence, observations of this diffuse emission and comparison with models is a fundamental tool for studying B-field, CR electrons and their propagation and distribution in the Galaxy.

It is known that there are at least two components of the Galactic B-field: regular and random \citep{rand}. 
However its intensity and configuration are still uncertain, despite large amounts of rotation measure (RM) data for pulsars and extragalactic sources \citep{TT, page, han2008, mao}. 
Information on the B-field  using synchrotron emission  came from early studies \citep{Higdon, haslam1981, Kanbach, beck2001}, followed by more recent ones \citep{strong2004, Kogut, sun2008, jaffe, jansson1, fauvet2011, fauvet2012}.
\cite{jaffe2010} used radio surveys of total synchrotron emission, polarized synchrotron
emission in the Galactic plane and RMs, and constrained the relative contributions
of the coherent, ordered random and isotropic random  components of the magnetic field, constructing a new model of some complexity. 
\cite{pshirkov} selected two benchmark models, fitting different B-fields to the observed RMs. 
\cite{jansson2}, instead,  
derived their best-fitting model including also a striated component. 
\cite{fauvet2012} produced 3D models of the Galactic magnetic field including regular and turbulent components only. 
Recently, \cite{Delabrouille} produced a model of the all-sky emission in the {\it Planck} satellite frequency ranges
with a regular component following a bi-symmetrical spiral formulation and a turbulent component. 

The additional component, in addition to the random and the regular B-fields, has been included by several authors in recent works\footnote{This is called ordered random by \cite{jaffe2010}, striated by \cite{jansson2} and anisotropic random by \cite{beck2013}. We call it here anisotropic random component to distinguish it from the pure random and regular components. One may adopt the name ordered to include both regular and anisotropic components, as in \cite{beck2013}.}. This anisotropic random component refers to a large scale ordering of the B-field, which originates by stretching or compression of the random field \citep{beck2001}. This component is expected to be aligned to the large scale regular field, with frequent reversal of its direction on small scales \citep{jaffe2010}. Radio observations of galaxies suggest that its intensity is stronger in the regions between the optical spiral arms \citep{beck2013}, due to the  shifting of the dynamo field  with respect to the density wave producing the spiral arms. However the proof of its existence is   model dependent and its structure is still not known. 
This component is not traced by RMs, but it contributes both to the polarized and unpolarized synchrotron emission.

Constraining  random and anisotropic random B-fields is possible using the combination of polarized and unpolarized observation of the synchrotron emission,  realistic CR electron spectra and distributions, and models of the B-field, combined with  RMs of pulsars and extragalactic sources.

In our previous paper \citep{strong2011}, hereafter SOJ2011, we focussed on the total synchrotron spectrum only at high Galactic latitudes, from which we derived information on the electron spectrum and the total B-field. 
We tested propagation models based on cosmic-ray and gamma-ray data against synchrotron data from 22 MHz to 94 GHz. 
We were thus able to put constraints on the total B-field intensity and propagation parameters.  
The synchrotron spectrum from this paper was used by \citet{Hinshaw2} as a foreground in deriving the final nine-year {\it WMAP} cosmological parameters.

Here, we start from the implementation of synchrotron emission in  GALPROP  as described in SOJ2011, and we extend the study to the whole Galaxy.
The novelty in this paper is the modelling and use of polarization, as well as the treatment of absorption and free-free emission. 
Spectral and spatial studies are performed.
For our spatial study and for the estimation of polarized emission, the 3D propagation scheme is necessary. 
This is the first time the 3D option has been extensively used with the GALPROP code (most previous works assumed radial dependence only, hence symmetry in x and y). 
In addition to the description of  the modelling improvements, 
in this paper we use GALPROP to calculate the radio emission of some illustrative models. 
The models are created using existing B-field models and realistic propagation models that are adjusted to match CR observations and gamma-ray data.  
We compare our prediction to radio data and investigate how a few selected parameters affect our results. 
We present Galactic profiles and skymaps, changing one parameter at a time, so that we fully understand the effects of different parameters and their degeneracy.  
We address the following three main questions: Do such  models resemble  the data? What can be learned from simple modifications of the parameters? How  degenerate are they? 
We use  recent CR electron and positron measurements by {\it Fermi}-Large Area Telescope (LAT) \citep{fermi_ele}
and  realistic CR source distributions in the Galaxy based on recent gamma-ray data analyses \citep{strong2010, diffuse2}.
No energy equipartition approximation between CR electron density and strength of the B-field is assumed.
 For the first time, models of both total and polarized synchrotron emission have been investigated in the context of CR propagation in a self consistent manner, with other recent observations such as gamma rays and CR measurements. 
For a given B-field and propagation model we start with a  CR source distribution and follow the propagation of all the
particles, primary protons, helium, electrons and positrons, taking into account
secondary production, and diffusion of particles and their energy losses throughout the whole Galaxy. 
We then compare the model results with CR measurements at earth and radio data. 
The contribution of secondary electron and positrons to synchrotron is  included.

The paper is organized as follows:
in the following section, we introduce new developments in the GALPROP code and give a  description of the implemented B-field models.
Then, we describe the method, present the results and discuss the effects of varying different model parameters.


\section{GALPROP developments}

GALPROP is a numerical code for modelling the propagation of cosmic rays in the Galaxy and calculating the diffuse emissions produced during their interaction with the interstellar medium (ISM) and radiation field (ISRF). 
Its objective is to assist in interpreting   observations of gamma rays, CRs, and theISM  in a self-consistent way. 
Description of the GALPROP software can be found in \citet{moska1998},  \citet{strong2004}, \cite{strong2007},  and the GALPROP Explanatory Supplement is available at the dedicated website\footnote{ http:$\slash \slash$galprop.stanford.edu } (see also  \citet{vladimirov} and references therein).
GALPROP is also used by the {\it Fermi}-LAT Collaboration for interpreting the observations of the Galactic interstellar gamma-ray emission \citep{diffuse2}. 
It simultaneously predicts CRs, gamma rays, and also synchrotron radiation \citep{strong2011}.
GALPROP solves the transport equation for all required CR species, given a CR  source distribution and boundary conditions. 
It takes into account diffusion, convection, energy losses and diffusive reacceleration processes. 
Secondary CRs produced by collisions in the ISM and decay of radiative isotopes are included.  
The propagation equation is solved numerically on a user-defined spatial grid in 2D or in 3D, and energy grid. 
The solution proceeds until a steady-state solution is obtained from the heaviest primary nucleus to electrons.
Positrons and anti-protons are then computed as well. 
GALPROP models gamma-ray emission of  pion decay, bremsstrahlung and inverse Compton for a user-defined CR source distribution and CR spectra. 
Gas maps and ISRF are provided and updated to the most recent observations \citep{porter2008, diffuse2}.

In this section, we describe the implementations on the GALPROP code for calculating both polarized and unpolarized synchrotron radiation in the radio band. 
We also include radio  absorption and a preliminary free-free emission model,  so that, with respect to SOJ2011, we are now able to model the emission also in the plane, where absorption at low frequencies is important. 
All the calculations reported here are performed with the 3D mode of GALPROP.
Given a spectrum of electrons or positrons computed at all points on the 3D grid and a B-field, 
GALPROP integrates
over particle energy to get the synchrotron emissivity.
The emissivity as seen by an observer at the solar position is
computed and output as a function of 
(x, y, z, $\nu$) (for the 3D case). The spectrum and distribution
of the emissivity depend on the form of the regular
and random components of the B-field, and the spectrum
and distribution of CR leptons. To obtain the synchrotron intensity
GALPROP integrates over the line-of-sight the calculated emissivity on the sky grid. 
The resulting synchrotron sky maps for a user-defined grid of frequencies are output in Galactic coordinates either as Cartesian projection (CAR) or in HEALPix.

\subsection{Synchrotron polarization}
We have implemented the Stokes parameters in GALPROP following the description for the  formulae as in the Hammurabi software \cite{hammurabi}. 
The emissivities for $I, Q$ and $U$ are defined by

\begin{equation}
\begin{array}{l}
\epsilon_I=\epsilon_\perp+\epsilon_\parallel \ , \  
\epsilon_P=\epsilon_\perp-\epsilon_\parallel \\ 
\epsilon_Q=\epsilon_P\cos(2\chi) \ , \ 
\epsilon_U=\epsilon_P\sin(2\chi)\\
\end{array}
\end{equation}

where $\epsilon_I$ is the total emissivity, $\epsilon_P = \sqrt{\epsilon_Q^2+\epsilon_U^2}$ is the polarized emissivity, $\epsilon_I-\epsilon_P = 2\epsilon_\parallel$ is the unpolarized emissivity. 
$\epsilon_\perp$ and $\epsilon_\parallel$ are the perpendicular and parallel components, relative to {\bf B}, of the synchrotron emissivity, as described in SOJ2011. 
$\chi$ is the angle between the polarization direction of the electric vector and the Galactic longitude meridian (N-S),
computed from the geometry of {\bf B} at each point. 
Unpolarized radiation has $\epsilon_Q=\epsilon_U=0$, totally polarized has $\epsilon_I=\epsilon_P$.

We integrate the emissivities  over the line of sight to produce the corresponding synchrotron sky maps of $I,Q$ and $U$; 
$P$ is computed from $P = \sqrt{Q^2+U^2}$, and the polarization angle from ${1\over2}\arctan(U/Q)$.
We use the same convention for $Q$ and $U$ as in \cite{page}.

Because the emission is polarized perpendicular to the projection of {\bf B} onto the line of sight,
the polarization vector is directed along $\chi+\pi/2$ 
 (N.B. $\epsilon_\perp> \epsilon_\parallel $ for synchrotron radiation).
With this definition of $\chi$,  {\bf B} parallel to the Galactic plane has $\chi=0,Q>0$ and $U=0$, while a projected  {\bf B} pointing towards the North Galactic Pole has $\chi=\pi/2, Q<0$ and $U=0$.  {\bf B} at $\pi/4$ to the meridian has  $\chi=\pi/4,Q=0$ and $U>0$, and  at $-\pi/4$ to the meridian  $\chi=-\pi/4, Q=0$ and $U<0$.

Note that the formulation here is more general than that often used, which gives the Stokes parameters directly in terms of the components of  {\bf B} (e.g., \cite{page}, {\it WMAP}). 
The latter is valid only under the assumption of a power-law CR electron spectrum with an index of 3, while ours is valid for any electron spectrum. 
The simplified formulae are useful for understanding the relation between the topology of  {\bf B} and the Stokes parameters, and for checking the software. 
These relations are as follows:

\begin{equation}
\begin{array}{l}
I\propto  \int ({\Bxp}^2+{\Bzp}^2)\ I_e(s)\ ds \\
Q\propto  \int ({\Bxp}^2-{\Bzp}^2)\ I_e(s)\ ds \\
U\propto  \int (2\Bxp \Bzp)\        I_e(s)\ ds
\end{array}
\end{equation}

where $\Bxp, \Bzp$ are the horizontal and vertical components respectively of  the projection of {\bf B}   on to  the plane  perpendicular to the line of sight to the observer,   $s$ is  the line-of-sight distance and $I_e(s)$ is the CR electron flux at $s$.  
The projection of  {\bf B} implies that $\Bzp$ is non-zero for latitudes $b\ne0$ even if there is no vertical component of {\bf B}; a consequence is that $U$ changes sign from positive to negative latitudes.

 Note that polarized emission is not sensitive to B-reversals, and hence neither is GALPROP.


\subsection{B-field representation}

In SOJ2011, we  extended GALPROP with a 3D description of the Galactic B-field, for regular
and random components. 
In our model of the Galactic B-field, $R$  is Galactocentric distance ($R=R_{\odot}=8.5$ kpc is the location of the Sun) and  $\theta$  is the azimuth angle measured anticlockwise from the Galactic centre-Sun direction. 
Our reference system is right-handed, with the Galactic centre at $x=y=z=0$ and the Sun in the $+x$ direction.

In the GALPROP coordinate system, the components  of the B-field projected on to the $x$- and $y$- axes  are
\begin{equation}
\begin{array}{rcl}
B_x & = & B_R \cos \theta - B_\theta \sin \theta\\
B_y & = & B_R \sin \theta + B_\theta \cos \theta\\
\end{array}
\end{equation}

The radial and azimuth components are defined as
\begin{equation}
\begin{array}{rcl}
B_R    & = & B(R,\theta, z)\sin p\\
B_\theta & = &- B(R,\theta,z)\cos p\\ 
\end{array}
\end{equation}
where $p$ is the pitch angle.


\subsection{Regular  B-field models}
\label{Bregular}

Our knowledge of the Galactic B-field is still very limited.
 The B-field intensity is measured using Zeeman spectral-line splitting, optical polarization data 
 and  Faraday RMs of pulsars and  extragalactic sources. 
RMs depend on the regular B-field component and the thermal electron density along the line of sight. 
The thermal electron  distribution is usually based on the  NE2001 model \citep{NE2001}, which is known be an inadequate approximation \citep{gaensler}.
 A complementary approach is to use  information from synchrotron emission both polarized and total.
 This is  dependent on the knowledge of CR spectra and distribution in the Galaxy. 
Another more physical approach is based on  Galactic dynamo models  \citep{beck2013, hanasz,brandenburg, gressel}. 

The regular B-field models used here are a few sample of simple models from the literature and based on RMs, where the analytical formulation is given in the original papers, so that it can easily be included in our code. 
For illustration, we show the three regular B-field models used in this work.


\begin{figure*}
\centering
\includegraphics[width=0.8\textwidth, angle=0] {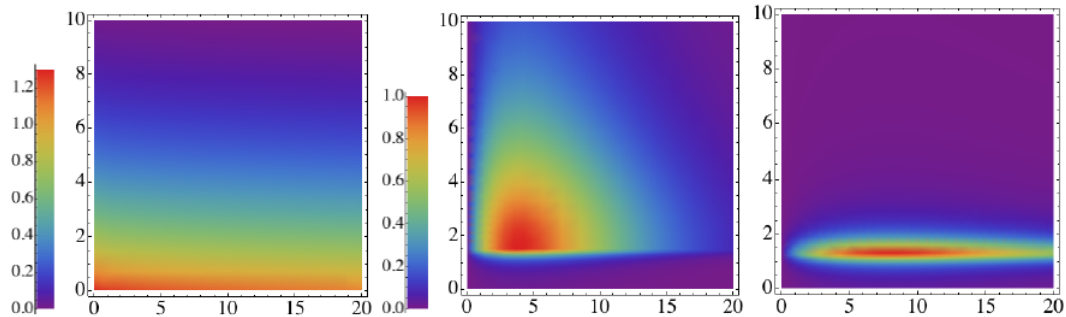}
\caption{Intensity of B($R,z$):  random component (left),  regular component of the halo for Model 1 (centre) and the regular component of the halo for Models 2 and 3 (right). Colors represent the intensity, with its maximum in the plane (for the random B-field) and at $z$=1.5 kpc (Model 1) and $z$=1.3 kpc (Model 2 and 3) for the regular halo field. Model intensities are shown for $B_{ran0} $ =  $B_0^H $ = 1$\mu$G. In this way, one can also obtain the all-sky intensity for all the models reported in Table \ref{Table2} by multiplying them by the best-fitting value $\approx$(4-7).  
 Horizontal axis marks the Galactocentric radius and vertical axis marks Galactic $z$. 
The models are symmetric in $z$, and hence only positive $z$ is shown, except that the halo field for Model 3 has $B_H(z<0)=-B_H(z>0)/2$. 
The regular disc field (not shown here) has an exponential form in $z$ with scalelength 1 kpc in all cases, and an $R$-dependence as shown in Fig~\ref{fig_disc}.}
\label{fig_halo}
\end{figure*}
\begin{figure*}
\centering
\includegraphics[width=0.7\textwidth, angle=0] {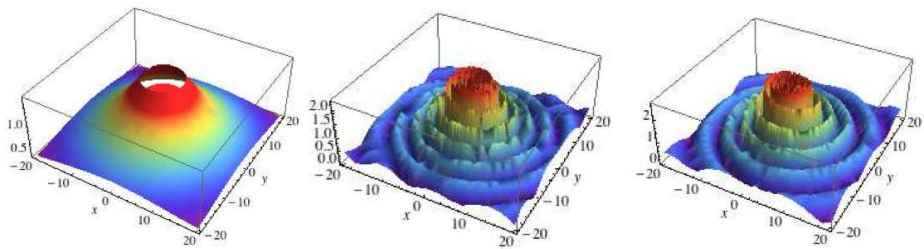}
\caption{ Intensity of the regular component of the disc B-field for Model 1 - 3 (left to right).
The graphics show the intensities as a function of the position in the Galactic plane ($z=0$),
with the sun at $x=8.5$ and $y=0$ kpc.
Model intensities are shown for a local B-field intensity of 1$\mu$G for all models. In this way, one can also obtain the all-sky intensity for all the models reported in Table \ref{Table2} by multiplying them by the best-fitting value $\approx$(3-4).}
\label{fig_disc}
\end{figure*}


\begin{itemize}
\item Model 1: as in \cite{sun2008} and \cite{sun2010} for disc and halo components. 
In their papers, the disc field was constrained by RMs near the Galactic
plane ($|b|<5\degr$) and the halo field by an all-sky compilation of RMs of extragalactic sources.
For the disc field, we took their axisymmetric spiral (ASS) model
plus reversals in rings (ASS+RING), which  best fitted their data.

\item Model 2 and Model 3:
as the logarithmic spiral models of the disc component from \cite{pshirkov} and their benchmark parameters.
After fitting different B-field models (based on Sun's models) to the latest set of RMs, they found that a spiral disk
and an asymmetric halo fits the data best, while the ring model is disfavoured.
We consider both their best-fitting ASS (defined here as Model 2) and bisymmetric spiral (BSS) (defined here as Model 3) models and their halo fields with their benchmark parameters.
Model 2 has pitch angle $-5\degr$, while Model 3 has  $-6\degr$. 
Model 2 has the same B-field intensity for the halo component in the Northern and Southern hemispheres, while in Model 3 the intensity in the Southern hemisphere it is half of the one in the north (which is the the same as in Model 2).  
\end{itemize}

For all these three disc models, the local B-field  intensity is the same, $B_0=2$~$\mu$G, as constrained by RMs.
 For the halo B-field, we used \cite{sun2010} for Model 1 and \cite{pshirkov} for Model 2 and 3 with their best-fitting parameters. Their halo field follows the double-torus
field as described in \cite{prouza2003}. 
Generating and maintaining a disc and a halo field of different parity in the same galaxy is under debate \citep{Moss2010,beck2013}.
Even if polarized emission is not sensitive to reversals, we used the models exactly as in the literature.
The models are illustrated in Figures~\ref{fig_halo} and \ref{fig_disc}. 
Note that they all have zero vertical component of the field.
For the regular disc component (Fig. \ref{fig_disc}), for the same B-field local intensity $B_0$, there corresponds a different maximum value for each model, which affects  the modelled synchrotron intensity distribution.

\subsection{Random B-field model}

The intensity of the random field is assumed to have the form
\begin{equation}
\label{eqBrand}
B_{ran}=B_{ran0}\,e^{ - (R-R_\odot)/R_B-|z|/z_B}.
\end{equation}
where the values of the parameters ($B_{ran0}, R_B, z_B$) can be freely chosen.
In SOJ2011 we were able to constrain the $\Bran$ field on the basis of the unpolarized synchrotron data, radio survey data
and the measurement of the electron spectrum by {\it Fermi}-LAT  \citep{fermi_ele}. 
The values of the parameters ($B_{ran0}, R_B, z_B$) found to fit
the 408 MHz synchrotron longitude and latitude profiles were $B_{ran0}= 7.5\mu$G for the local B-field and the  scalelengths were 30~kpc in $R$ and 4~kpc in $z$. 
These values were obtained adopting the \cite{sun2008} and \cite{sun2010} models for the regular B-field, with no attempt to fit the polarization data. 
In fact the original \cite{sun2008} value of  $B_{ran0}= 3\mu$G everywhere underestimated the total synchrotron emission. The shape of the random B-field is chosen to better resemble the intensity of the Galactic synchrotron emission as already introduced in SOJ2011.

\subsection{Anisotropic random B-field model}
We have also included  an anisotropic random component, which follows the same topology of the regular field, as in \cite{jansson2}.
We assume here the anisotropic random
component to be aligned with the local regular field but changing direction randomly on small scales.
This component  contributes to the polarized and total synchrotron emission, but not to the RMs.  

\subsection{Free-free absorption and emission model}
Synchrotron emission is affected by absorption by ionized hydrogen.
This is important at lower frequencies, but it also slightly affects the 408 MHz data in the Galactic plane. 
Hence, in our computation with GALPROP, we have also introduced a model for absorption. 
As a result, we can produce a synchrotron model for the whole sky and compare it with data for each frequency used. 
We use the formulae given in \cite{allen} (Chapter 5.9) for the free-free opacity (k$_{ff}$) and emissivity (e$_{ff}$), where
\begin{equation}
\begin{array}{rcl}
k_{ff} (\nu, N_e, T_e) & = & 0.0178 ~g_{ff}(\nu, T_e)~ N_e^2 /(\nu^2 T_e^{3/2})\\
e_{ff} (\nu, N_e, T_e) & = & 5.444~e^{-39} ~g_{ff}(\nu, T_e)~ N_e^2 /( T_e^{1/2})\\
\end{array}
\end{equation}
with g$_{ff}$($\nu$, T$_e$)=10.6 ~+~1.9~log(T$_e$)~-~1.26~log$_{10}$($\nu$).

Absorption is a function of the electron temperature T$_e$ and the clumping factor since it depends on the square of the electron density. 
The clumping factor is related to the filling factor: for equal-size clouds, clumping factor = 1/filling factor. 
Clumping factors  in the range  10-100 are typically based on a variety of data (pulsar dispersion measures, H$\alpha$ emission, free-free emission and synchrotron absorption). 
Estimates for  T$_e$ vary between 1000 and 9000 K \citep{gaensler}.

Note that free-free absorption follows the  same function of the temperature and the clumping factor as free-free emission. 
Free-free emission contributes to unpolarized radio emission at frequencies above few GHz in the Galactic plane. 
Hence, in order to model the total radio emission, a template of free-free emission is necessary. 
Even though available to estimate this emission component, the {\it WMAP} free-free template is not suitable, because it overestimates the free-free emission, presumably due to contamination by the anomalous dust emission, as pointed out by \cite{Alves}.

We have  introduced  our preliminary model for free-free emission in GALPROP, which will be constrained together with the synchrotron emission itself in future studies. 
The free electron density  implemented in GALPROP uses the model  described in \cite{cordes} and \citep{NE2001},
with the broad ionized component updated according to \cite{gaensler}.
We use the smooth part of the model only and do not model the spiral structure aspects.
Future developments should include  a more sophisticated model of the warm ionized medium, e.g.  \cite{Schnitzeler}.
 In this paper an electron temperature of 7000 K and a clumping factor of 100 are assumed. 
This clumping factor was chosen so that the model  resembles the total {\it WMAP} intensity in the inner Galaxy, in the presence of the other emission components.
We use the free-free emission model for completeness and illustration only. 
No attempt is made to study in detail its parameters, and our conclusions are not affected by different choice of the parameters.
Note also that free-free emission is subject to absorption as well as the synchrotron emission, and this is included in our model.
 

\section{Observational data}
Radio surveys used throughout this paper are extensively described in SOJ2011. 
We provide here a brief summary. 
We use the full-sky 408 MHz survey (Haslam et al. 1982), 
after correcting it for CMB, extragalactic non-thermal component and zero-level by subtracting 3.7K, as found by \cite{Reich1988}.
Other surveys were obtained directly from their authors, and others from the Bonn\footnote{http://www.mpifr-bonn.mpg.de/survey.html} and LAMBDA\footnote{http://lambda.gsfc.nasa.gov} websites. The combined zero-level and
extragalactic/CMB corrections were taken from the literature as stated here:
22~MHz - Dominion Radio Astrophysical Observatory Northern hemisphere survey: \citet{1999A&AS..137....7R}. 
45~MHz - north: \citet{1999A&AS..140..145M}
, south: \citet{1997A&AS..124..315A}
, combined all sky: \citet{Guzman2011}. 
 An offset of 550K was subtracted  \citep{Guzman2011}.
 150~MHz - Parkes-Jodrell Bank all-sky survey: \citet{1970AuJPA..16....1L}. 
 1420~MHz - Stockert-Villa Eliza all-sky survey. North: \citet{1982A&AS...48..219R,1986A&AS...63..205R},  
 south: \citet{2001A&A...376..861R}. 
An offset~of 2.8K was subtracted \citep{Reich2004}.
 2326 MHz - Rhodes southern hemisphere survey:  \citet{1998MNRAS.297..977J}. 

In addition to these data,
total and polarized data at 23, 33, 41, 61 and 94 GHz  {\it WMAP} seven-year data were obtained 
from the LAMBDA website \footnote{http://lambda.gsfc.nasa.gov/} \citep{gold}; see the Appendix for the identification and properties of the datasets actually used here.
The {\it WMAP} MCMC templates are used for dust and spinning dust. 
The  \citet{miville} total synchrotron maps are used to compare our model to the total synchrotron.
 The {\it WMAP} synchrotron templates were used as a cross-check.

We use $P$  to constrain the magnitude of the ordered (regular plus anisotropic random) B-field\footnote
{Noise bias in $P$ induced by noise in $Q$ and $U$  is addressed in the Appendix.};
however, separation into $Q$ and $U$ Stokes parameters is also presented,
since this is  useful to probe the field topology.
Above 23~GHz, the polarized emission has both synchrotron and thermal dust contributions.
 Hence, for completeness we include also the template of polarized dust from {\it WMAP} in our plots.
 However, we use  only 23~GHz  to draw conclusions on the polarized synchrotron emission, since at this frequency the polarized dust emission is much less than the synchrotron. 

 The total  $I$ also requires the inclusion of other components: free-free, dust and spinning dust. 
Hence, {\it WMAP} templates for the dust and spinning-dust emission are  used, since they are important especially in the Galactic plane. 
We use $I$ in the {\it WMAP} frequency ranges for illustration only, since it is not used to draw conclusions on the B-field.

For our spatial analysis we focus mainly on  the  408~MHz and 23~GHz data;
the 408~MHz map traces the total B-field and reflects mainly the synchrotron component, while the free-free emission is a minor component.
 The 23~GHz polarized map traces the synchrotron emission  related to the ordered (regular plus anisotropic random) B-field,  while the polarized dust component is minor.

\section{Testing existing B-field models}

We start by using the plain diffusion model of CR propagation as in SOJ2011. 
This propagation model describes both CR, gamma-ray and synchrotron spectral data \citep{strong2010, orlando, orlando2013}.
The initial model has a halo height of 4 kpc, and is based on injected spectra and CR source distribution from \cite{strong2010}.
CR injection spectra for nuclei and primary electrons are the same as those used in \cite{strong2010} for their plain diffusion model (LMPDS), as given in the Supplementary Material of that paper.

\begin{figure}
\centering
\includegraphics[width=0.3\textwidth, angle=0] {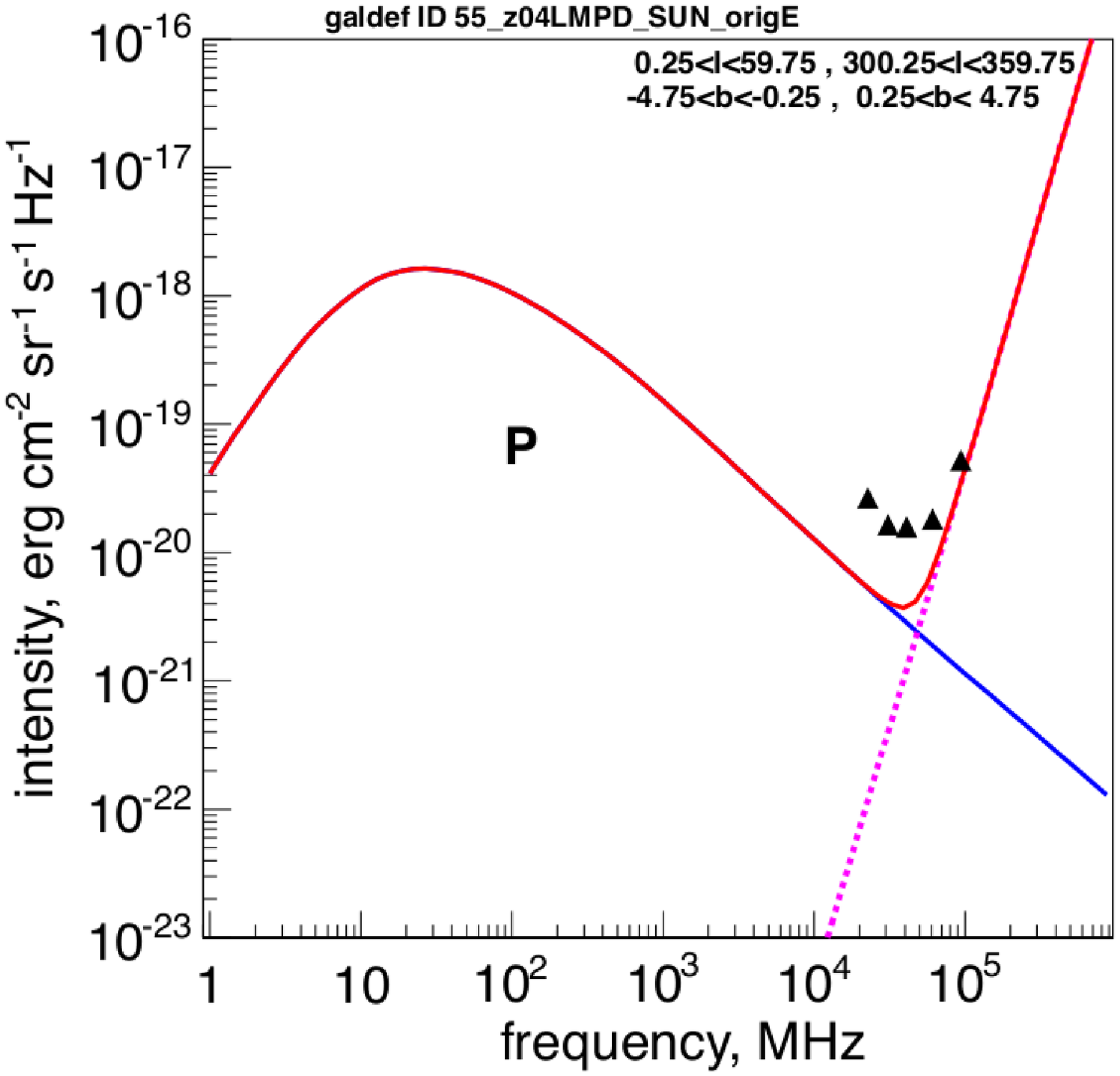}
\includegraphics[width=0.3\textwidth, angle=0] {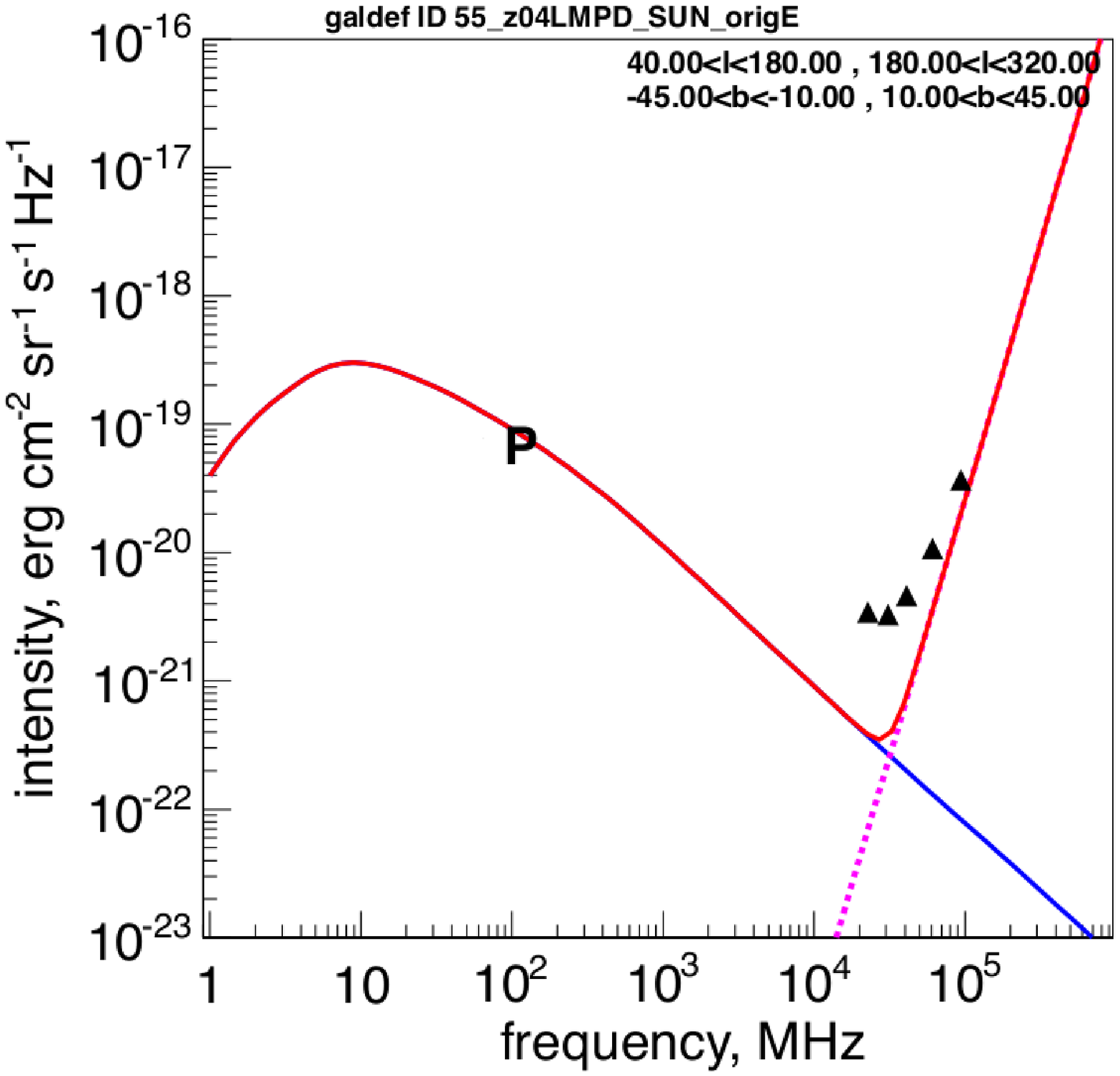}
\caption{Model 1:  spectra P for the inner Galaxy (upper plot) and for high latitudes (lower plot) and z=4 kpc. Regular B-field intensities are the same as the original models. The plots show the different model components: synchrotron (blue line), thermal dust (pink dotted line). Data are from {\it WMAP} (see details in the text). }
\label{fig_sun_original}
\end{figure}

For the original models described in Section \ref{Bregular} with intensities of the regular B-field fixed to the values given in the cited publications, our computed synchrotron emission does not match the polarized data at 23 GHz, where the emission can be considered totally synchrotron.
In fact, the expected emission is largely underestimated for all three B-field models described above. 
Figure \ref{fig_sun_original} shows as an example, the polarized spectra for Model 1 for the inner Galaxy and high latitudes. 

The energies of CR electrons responsible for synchrotron emission at that frequency are essentially  exempt from the effect of solar modulation. 
Hence, assuming that CR electrons throughout the Galaxy are reasonably modelled by GALPROP, this provides evidence for the presence of  an additional synchrotron component. 
We attribute this emission to be produced by an anisotropic random component, in agreement with recent work by \cite{jaffe} and \cite{jansson2}. 

The electron spectrum used  by \cite{sun2008} and  \cite{sun2010} is a factor of 2 higher than the one used in our analysis.
 This could partly explain why they obtained agreement with the polarized synchrotron data with no inclusion of the anisotropic random B-field. 
Note that, the inclusion of an additional component of the ordered (regular plus anisotropic random) B-field is tested in the 23 GHz map, where solar modulation is negligible, so the electron spectrum is known from direct measurements. 

We investigate the contribution of the anisotropic random component and quantify it in Section \ref{results}.


\section{Fitting procedure} \label{fitting procedure}

We consider the intensity of the regular B-field fixed to the values reported in the respective papers.
No attempt is made to scan over all the parameters or to construct new B-field models.

Initially, we compute each model with reference values for $B_0\ aniso$, $B_0^H\ aniso$ and $B_{ran0}$ 
derived by approximate eye fits
to polarized spectra in the inner Galaxy, where the P-bias is lower  (see the appendix for discussion of P-bias). 
The intensity of the halo field is adjusted maintaining the original scaling factor between the disc and the halo components of the regular B-field.
The initial random B-field is then adjusted to match the 408 MHz map in the plane. 

After a first approximation given by eye fitting, we started the fitting as follows.
Although, the fitting could be performed by using multiple GALPROP runs with varying parameters, this is not necessary,
and a simpler approach is fast and flexible, giving the same results.
The sum of the regular and anisotropic random field scaling is determined using $P$ at 23 GHz. 
Then, the random field scaling is determined using $I$ at 408 MHz, accounting
for the part $I_{reg+aniso}$  attributed to the fitted regular plus anisotropic random B-field (polarized fraction from regular field $0.70-0.75$). 
The free-free contribution at 408 MHz (around 10\% in the inner Galaxy) is also subtracted from the $I$ data before fitting.
An offset term accounts for zero level uncertainties,
 and P-bias in the case of polarized intensity.
 Also for the 408 MHz data an offset has been fit.
After the intensity normalization for $P$ and $I$ has been fitted, 
the corresponding B-field is obtained using the fact that $P\propto B_0^2$,  $I-I_{reg+aniso}-I_{ff}\propto B_{ran0}^2$;
the $B^2$ relation is  a good approximation for an electron spectral index 3, which is the case for electron energies ($>$10 GeV) producing the synchrotron at these frequencies.
We then rerun the models with the best-fitting values to provide a check on the procedure, with further fitting iterations if required.

We use a simple $\chi^2$ minimization, in order to make the fit and obtain the best values.
This is done on a HealPix grid, which has uniform solid angle bins not requiring any weighting.
The measurement error on the data points is assumed constant.
For the 408 MHz analysis, the data were sampled on the GALPROP model order-6 HealPix grid, with 49152 data points, resolution $\approx 1^o$.
For {\it WMAP}, the number of data points is  3145728,
corresponding to the order-9 HealPix (pixel size 6\arcmin) data provided by {\it WMAP}. Since we use the $1^o$ smoothed {\it WMAP} data,
 $\chi^2$ is effectively scaled up by a constant relative to independent pixels, but this was considered more reliable than resampling on the GALPROP model grid.
The whole sky is included in the fitting.
The absolute value of  $\chi^2$ is arbitrary, but can be used to compare models\footnote
{Since the real sky has many complex structures which our models do not even attempt to reproduce, a formal quality assessment has no meaning,
while model comparison is nevertheless meaningful.
 For example,  Loop I  (the North Polar Spur) and other `loops and spurs'  are not included in our modelling, since they are believed to be local structures and need further dedicated modelling. }.

\begin{table*}
\begin{center}
\caption{Summary of model-fitting results for three different regular B-field models.
 The values of $B_0$ (regular), $B_0^H  $ and $B_{ran0}$  are derived from the
fitted  normalizations as described in the text. All the models in this table uses the plain diffusion propagation model with halo height = 4 kpc and CR source distribution by \citep{strong2010}. See the supplementary material of their paper for more details.}
\begin{tabular}{lccc}
\hline
\\       
  Model code & SUNE & PSASSE  &   PSBSSE \\
 \hline
\\                       
    Figure & \ref{SUN_spectra}+\ref{SUN_prof} & \ref{ASS_prof}& \ref{BSS_prof}\\
 \\
{\bf P(23 GHz)}\\
     B model &      1 &       2 &       3          \\
    $B_0\ regular $ ($\mu$G)  &    2    &       2                    &   2               \\
    $B_0^H  $ ($\mu$G)        &    2    &  4 (North)  ~~  2 (South)   & 4 \\
     $B_0\ aniso $ ($\mu$G)  &    2.28    &       1.78                    &   1.09               \\
    $B_0^H\ aniso   $ ($\mu$G)        &    2.28    &  3.56 (North)  ~~  1.78 (South)   & 2.18 \\
     Pitch angle, $p$ (deg)&-12& -5 &-6  \\
     offset  ($mK$)& 0.0094 &  0.013     & 0.012                     \\
     rms error  ($mK$)&0.033  & 0.033     & 0.033                         \\
    $\chi^2$ &      3430  &     3455 &       3501      \\
    \\
{\bf I(408 MHz)}\\
   $B_{ran0} $ ($\mu$G)  &   5.16 &  5.38   &  5.38      \\    
   offset  ($K$)         &4.71  & 5.2   &4.85               \\
    rms error  ($K$)&10.7 & 11.0    & 10.9              \\
   $\chi^2$ & 3.65e+08  &3.82e+08  &  3.76e+08       \\
\\

 \hline
 \hline
\label{Table1}
\end{tabular}
\end{center}
\end{table*}

\begin{table*}
\begin{center}
\caption{Summary of figures and model parameters. }
\begin{tabular}{lccccccc}
 \hline
 \hline
\\
         Model code & SUNE  & SUN10E& SUNLorimE &SUNLorimv33E\\                   
  \\
 \hline
 \hline
\\
{\bf Propagation model }\\
  \\
  Halo size, $z$ (kpc) & 4 &10&4&4\\
  \\
  Source distribution$^{a}$& S2010&S2010&L2006&L2006\\
  \\
  Reacceleration, $v_A$ (km s$^{-1}$) &0  &0&0&33\\
\\
B model &      1 &       1 &       1&1          \\
         \\
 \hline
\\
{\bf Fixed B-field intensity}$^{b}$\\
  \\
      Figure & \ref{SUN_spectra}+\ref{SUN_prof} & \ref{SUN10_prof} & \ref{SUNLorim_prof} & - \\    
        \\
 \hline
\\
{\bf Fitted B-field intensity}$^{c}$\\
    \\
\\
P(23 GHz)\\
 ~ ~   $B_0\ regular $ ($\mu$G)  &    2              &   2   & 2      &   2     \\
    ~ ~$B_0^H  $ ($\mu$G)        &    2    &2& 2      &   2     \\
    ~ ~   $B_0\ aniso $ ($\mu$G)  &    2.28              &   2.11   & 1.74      &   1.46     \\
    ~ ~$B_0^H\ aniso $ ($\mu$G)        &    2.28    &2.11& 1.74      &   1.46     \\
   ~ ~  offset  ($mK$)& 0.0094   &  0.008         &  0.013 &  0.012                 \\
    ~ ~ rms error  ($mK$)&0.033        &  0.034        &  0.034 & 0.034                  \\
   ~ ~ $\chi^2$ &      3430 &  3671 &    3626  &    3711      \\
  \\
 lower frequencies.


 I(408 MHz)\\
~ ~   $B_{ran0} $ ($\mu$G)  &   5.16        &5.03 &  4.74       & 3.27     \\   
 offset  ($K$)         &4.71       &  0.98         &8.1 & 6.2                 \\
  ~ ~  rms error  ($K$)&10.7       &  11.8         & 11.8 &  12.3               \\
 ~ ~  $\chi^2$ &  3.65e+08 &4.39e+08&  4.36e+08             &      4.78e+08         \\
\\
 \hline
 \hline
\label{Table2}
\end{tabular}
\end{center}
$^{a}$For all the models $B_0\ regular $ = $B_0^H  $ = 4.28 $\mu$G and $B_{ran0} $ = 5.3 $\mu$G, best-fitting values for SUNE. \\
$^{b}$S2010 = \cite{strong2010}, L2006 = \cite{lorimer}.\\
$^{c}$These parameters are used for  the sky maps.
\end{table*}


\section{Results}
\label{results}

We first show the results of the three B-field models and the best-fitting parameters  following the fitting procedure described in Section \ref{fitting procedure}. 
We then choose the model that gives the best fitting to the data and use it as the sample for testing  different propagation models.
To test the effects of the propagation model, we vary some of the parameters, namely the propagation halo size and CR source distribution. We also test the spatial effect of reacceleration.
In order to separate the effects of CR propagation from the B-field model, we keep the propagation model fixed while varying the B-field model, and then we keep the B-field model fixed while varying the propagation model.

\subsection{Effects of different B-field models }

To study the effect of different B-field models, we fix the propagation parameters while we vary the B-field.
The plain diffusion propagation model, a halo height of 4 kpc and CR source distribution from \cite{strong2010} are used. 
The results of the model fitting are summarized in Table~\ref{Table1}.

\begin{figure*}
\centering
\includegraphics[width=0.4\textwidth, angle=0] {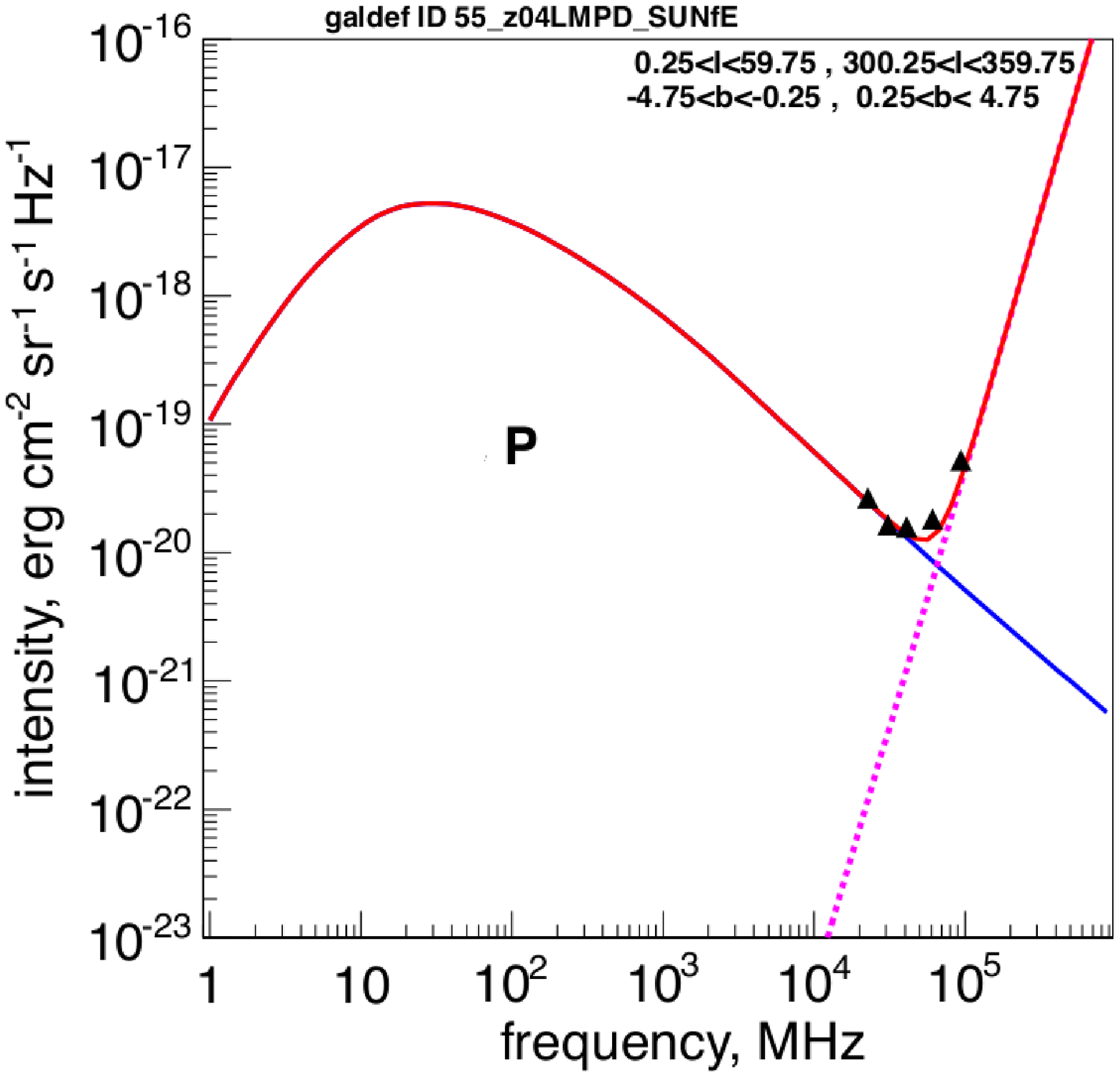}
\includegraphics[width=0.4\textwidth, angle=0] {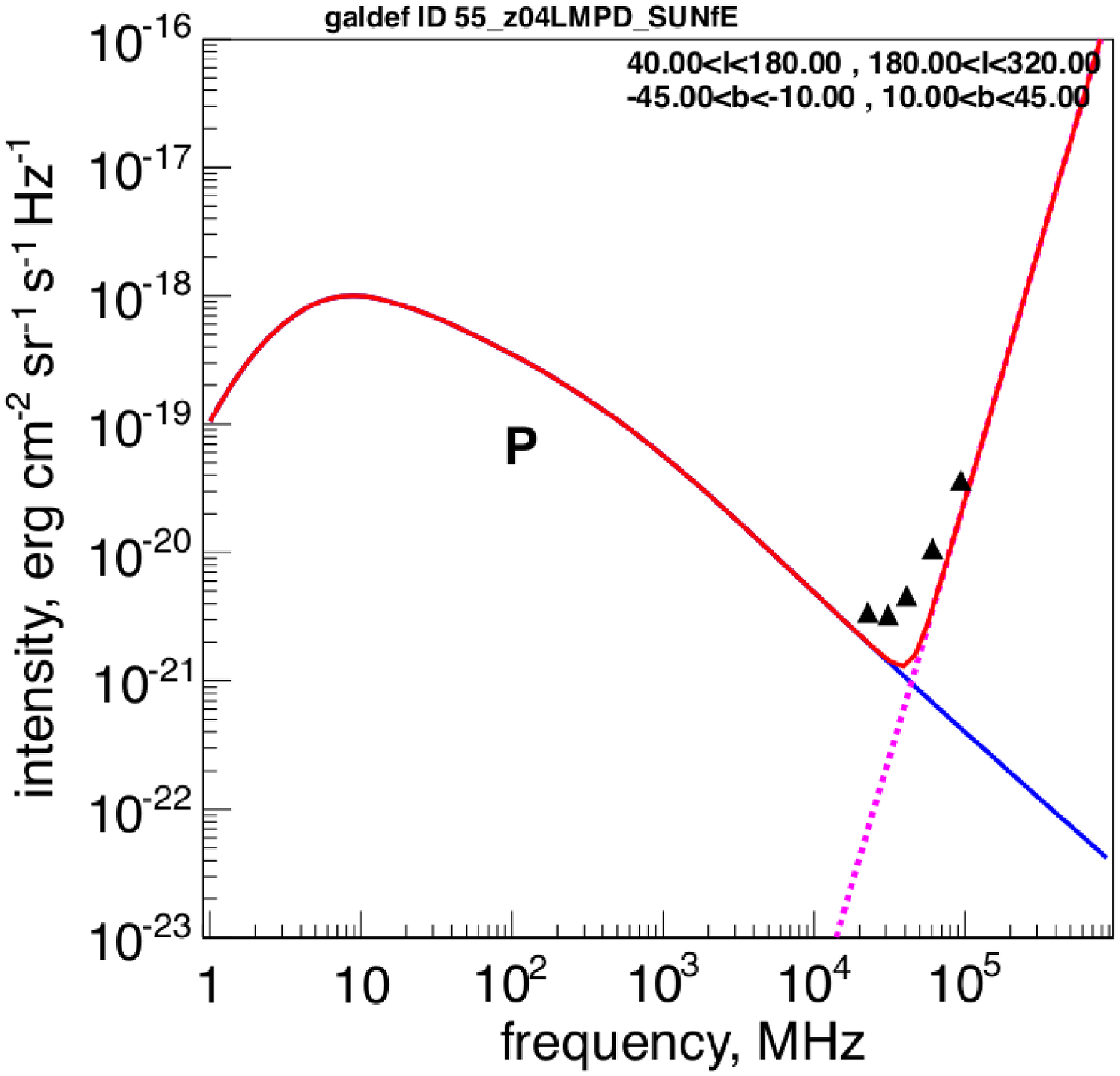}
\includegraphics[width=0.4\textwidth, angle=0] {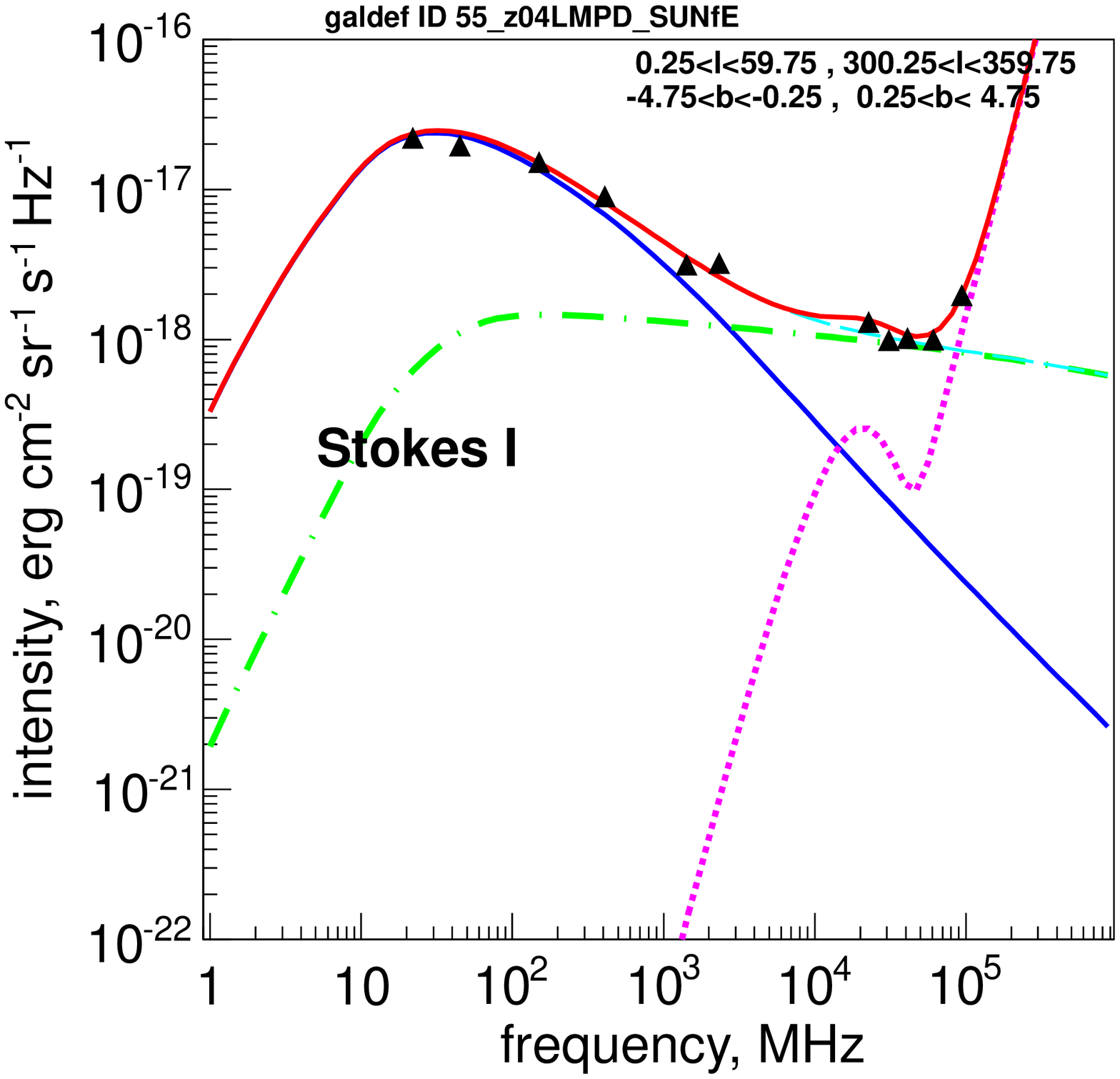}
\includegraphics[width=0.4\textwidth, angle=0] {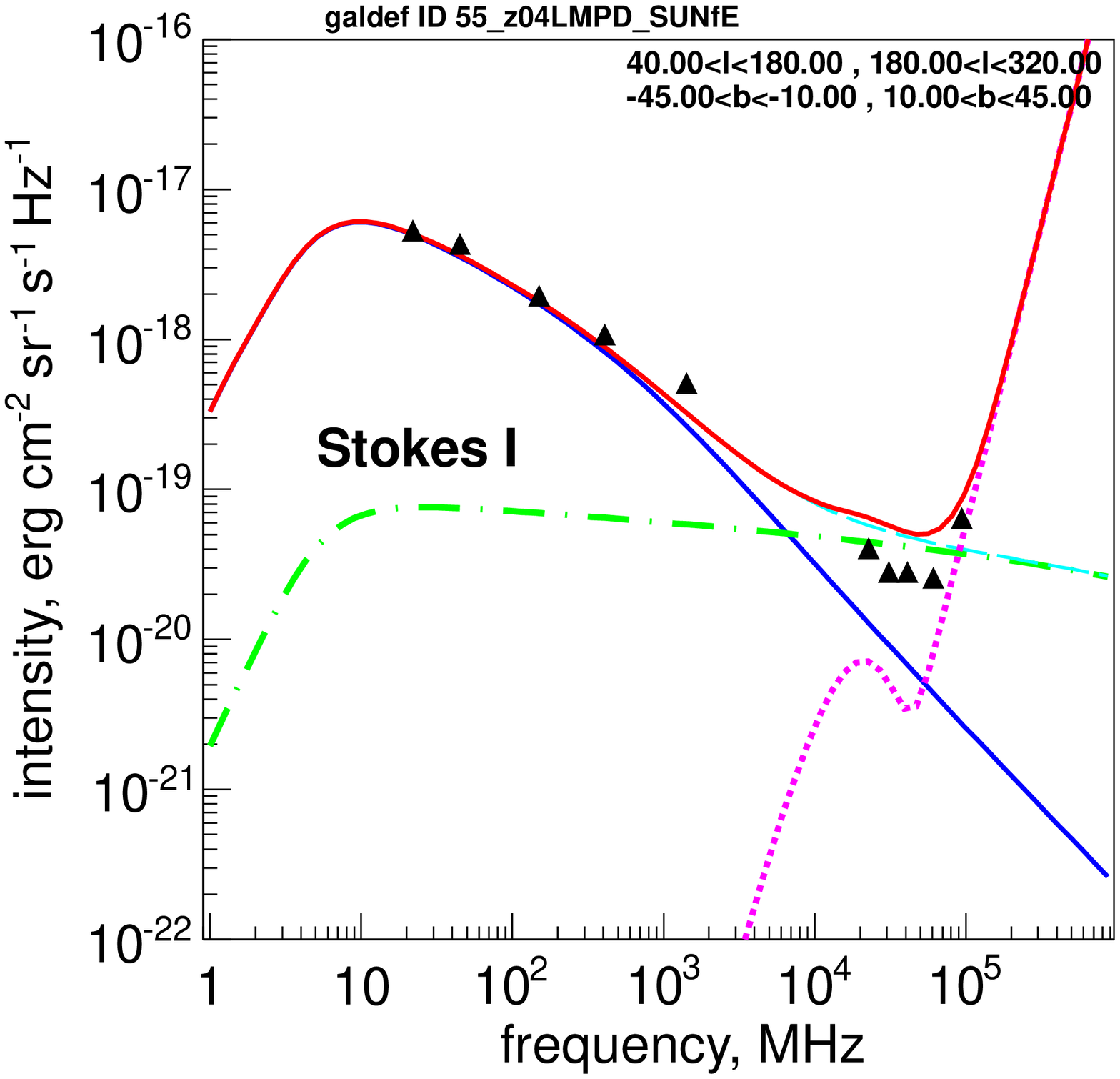}
\includegraphics[width=0.4\textwidth, angle=0] {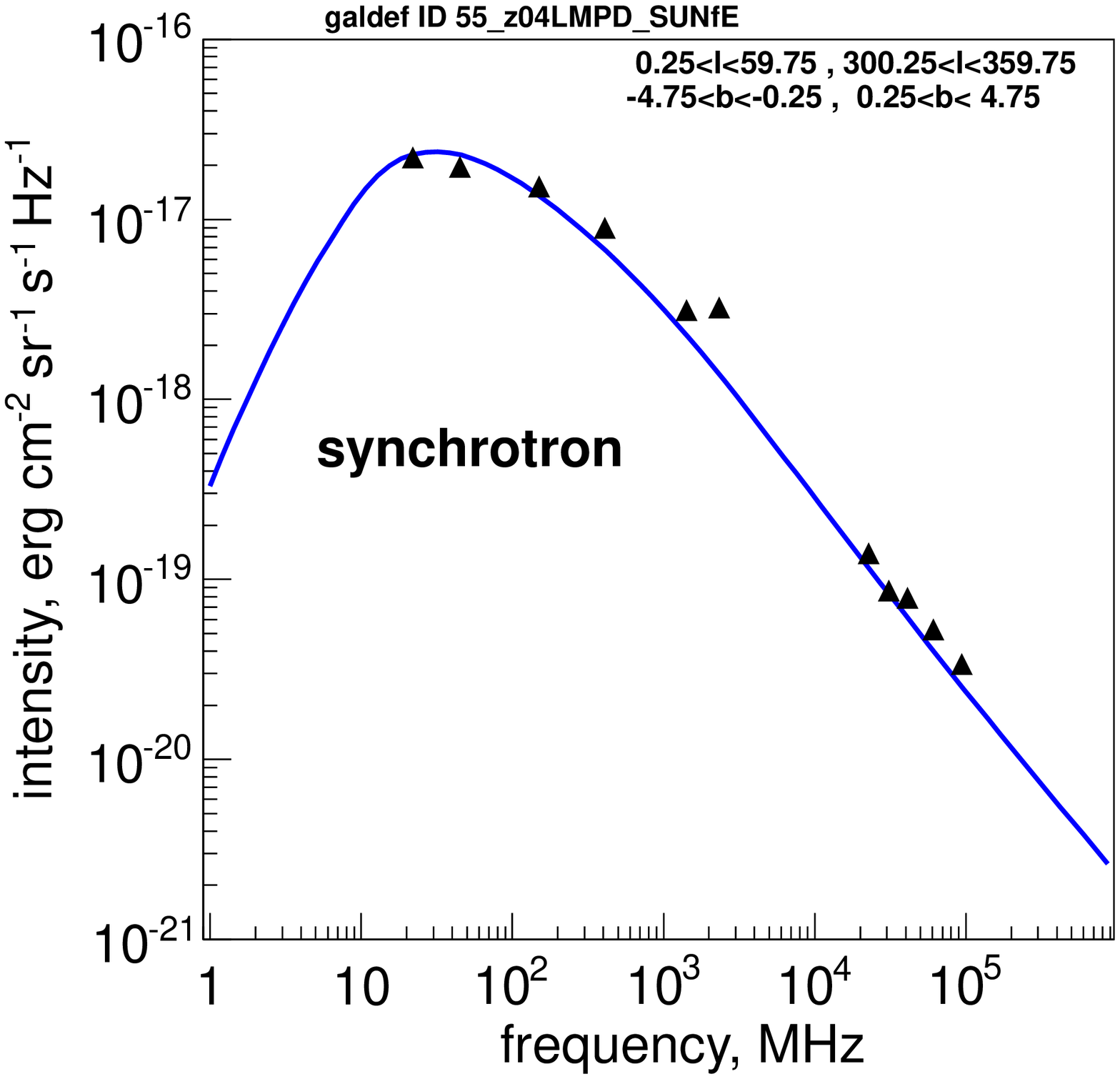}
\includegraphics[width=0.4\textwidth, angle=0] {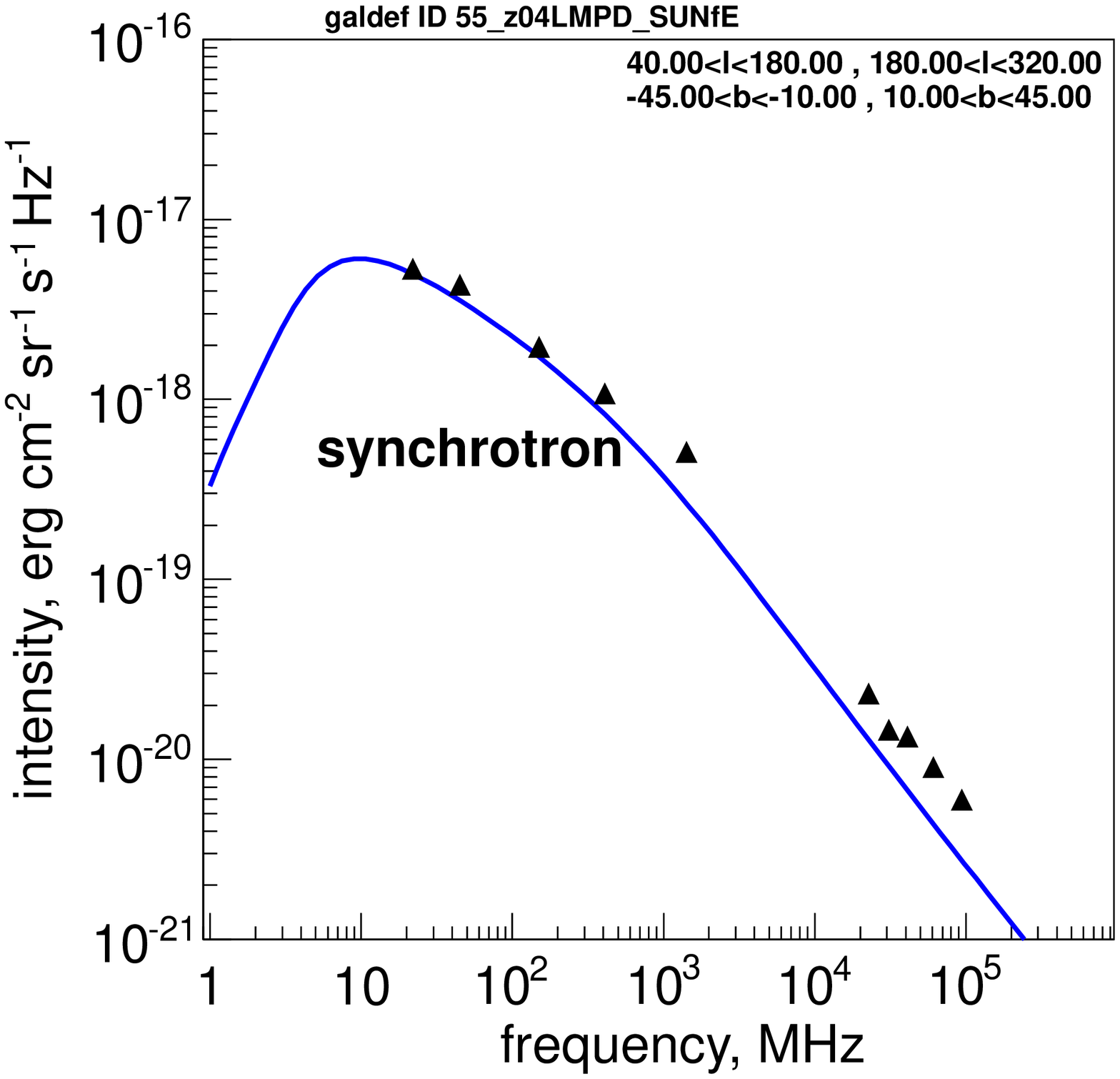}
\caption{  Model SUNE:  spectra $P$ and $I$ and only-synchrotron for the inner Galaxy (left-hand plots) and for high latitudes (right-hand plots) and z=4 kpc. B-field intensities are scaled with respect to the original models to agree with data. The plots show the different model components: synchrotron (blue line), dust and spinning dust (pink dotted line), free-free (green dashed-dotted line), free-free plus synchrotron (cyan dashed line) and total (red line). Data (black triangles) are from radio surveys  and {\it WMAP} (see details in the text).}
\label{SUN_spectra}
\end{figure*}

Figure \ref{SUN_spectra} shows the polarized, total and synchrotron spectra compared with data for two sky regions: the inner Galactic plane and high latitudes (avoiding the  North Polar Spur). 
Spectra are shown  for the first B-field  model ($SUNE$) only, 
 since they are nearly identical for the same electron spectral index and propagation model. 
For the inner Galaxy spectrum, the  predictions  are in agreement with data for $P$, $I$ and the synchrotron-only component for all the models,
while at high latitudes $P$ and the synchrotron-only component predictions are underestimated, which suggests that the B-field is not properly modelled over the whole Galaxy.
By contrast, the total intensity $I$ in the {\it WMAP} range is overestimated. 
This suggests that the  free-free emission component is not well modelled at high latitudes.
For all regions, the total spectrum at lower frequencies is well fitted, which indicates that the electron local spectrum is well modelled, as in SOJ2011.
 
\begin{figure*}
\centering
\includegraphics[width=0.3\textwidth, angle=0] {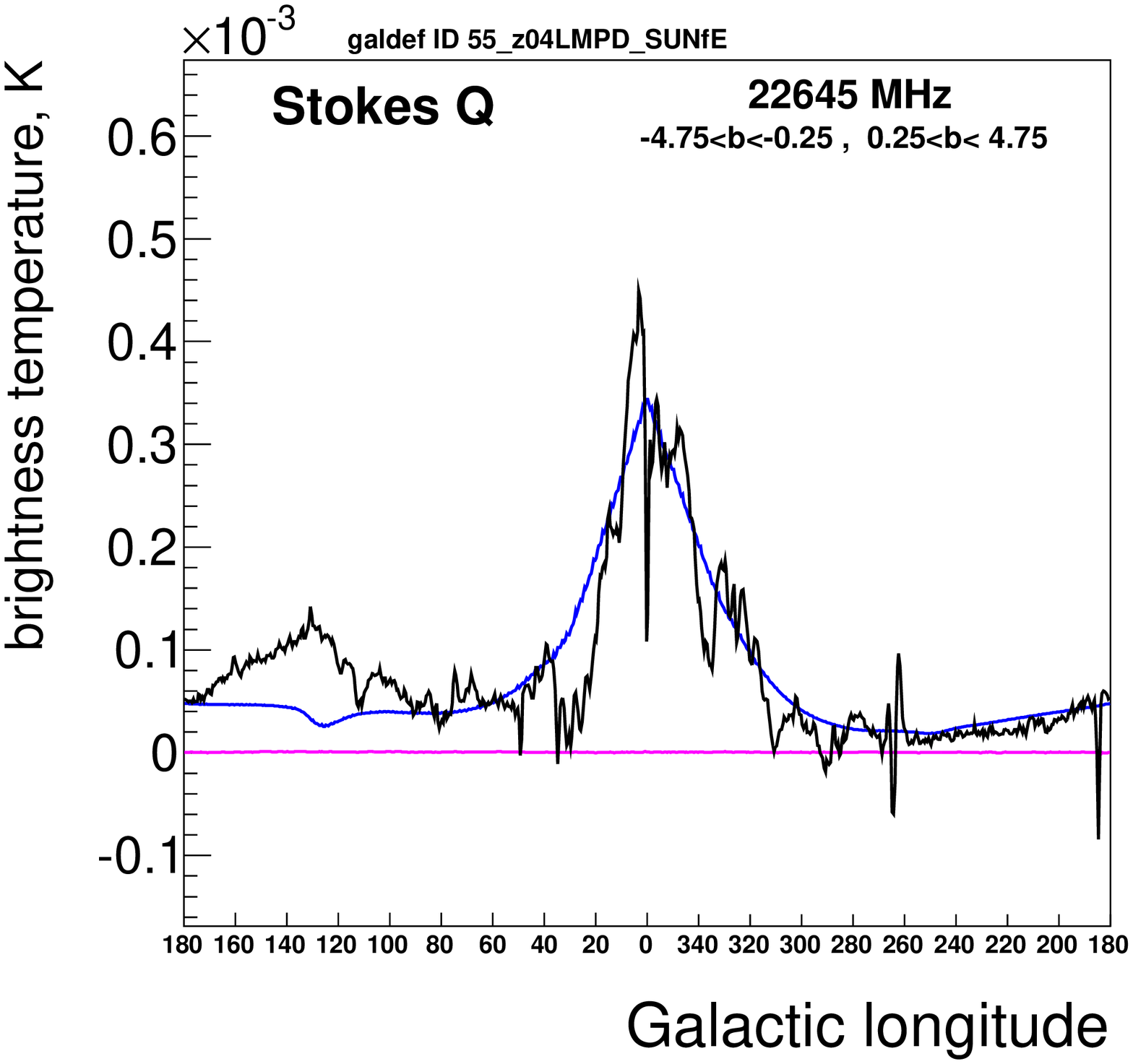}
\includegraphics[width=0.3\textwidth, angle=0] {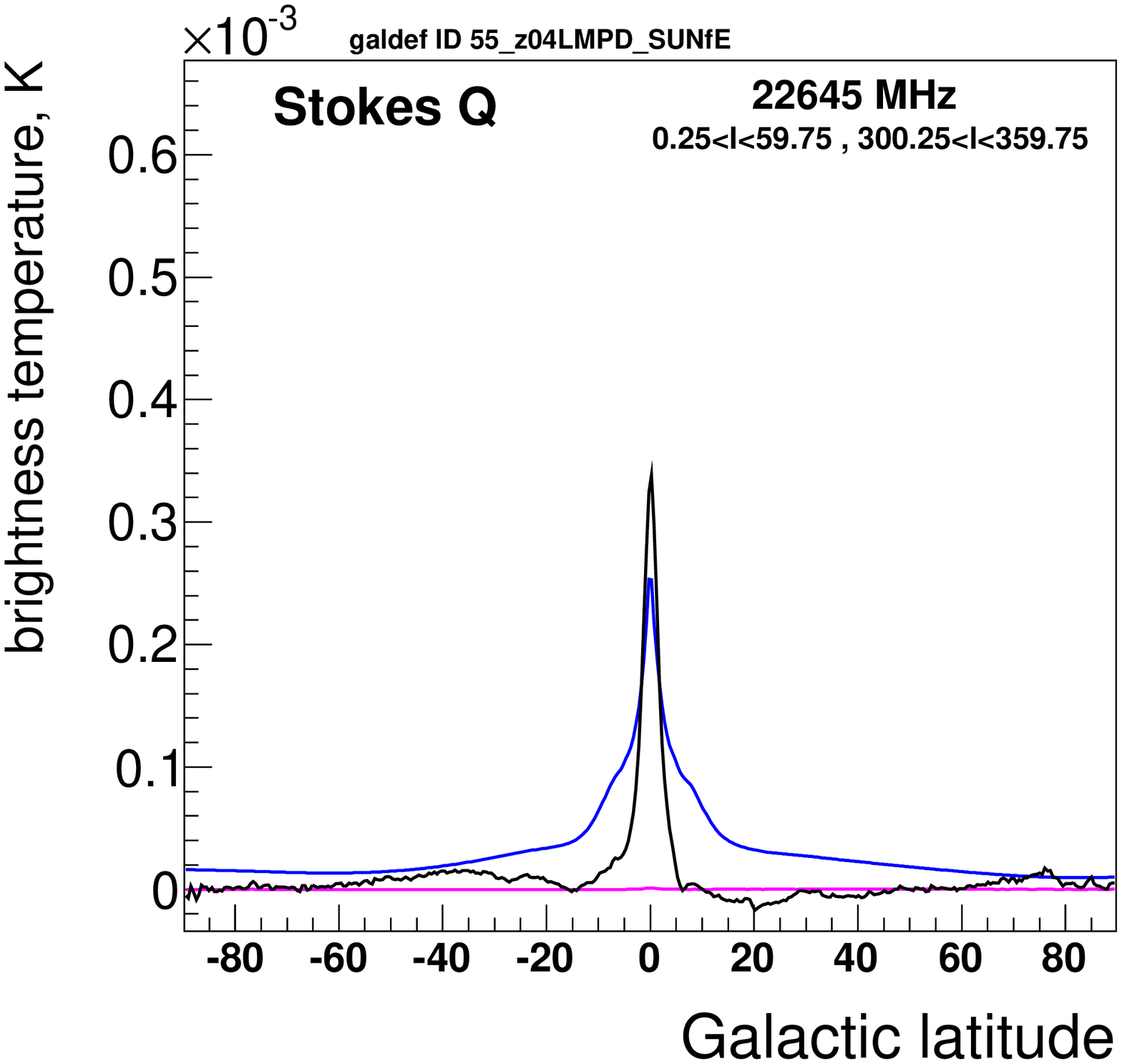}
\includegraphics[width=0.3\textwidth, angle=0] {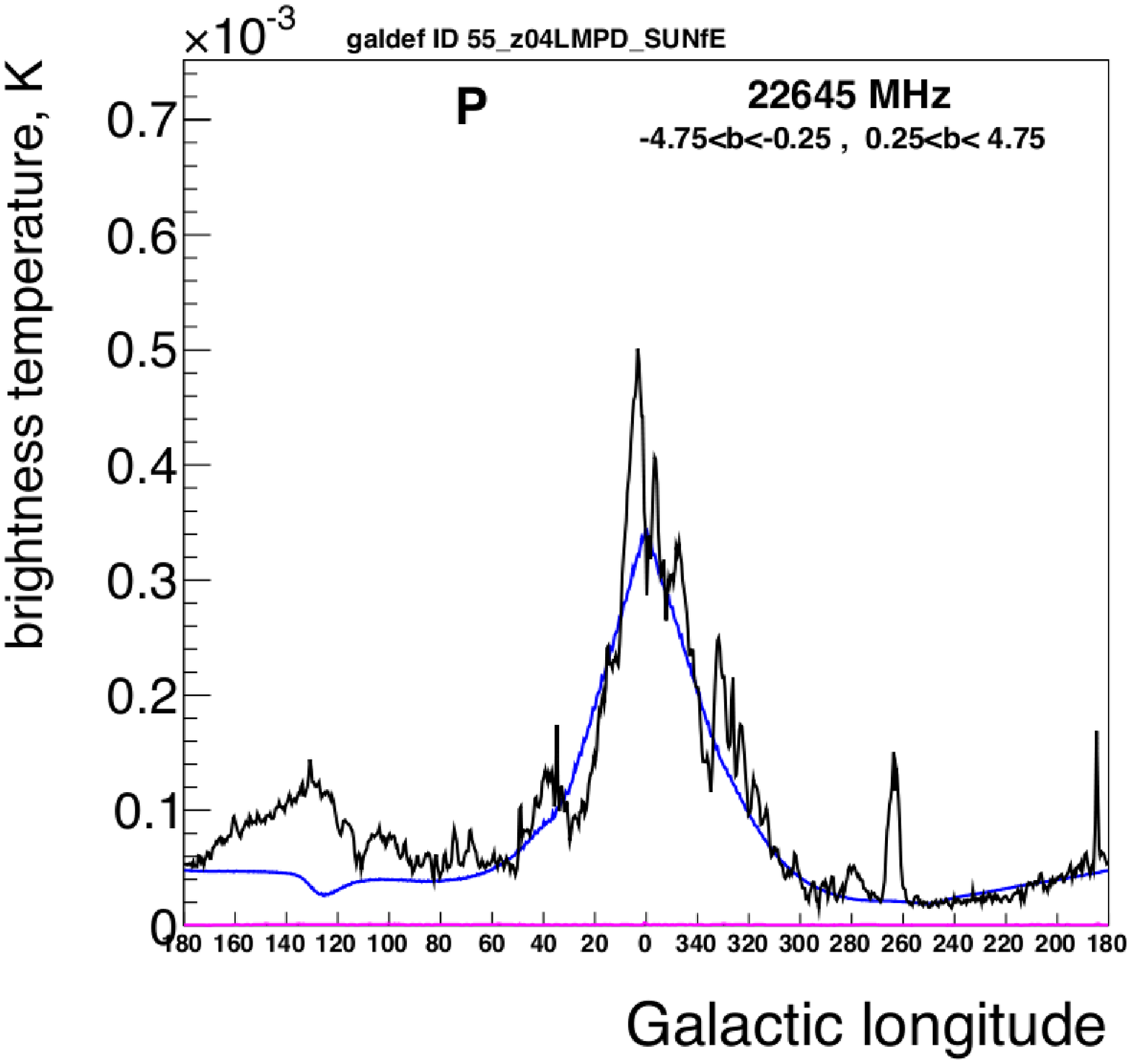}
\includegraphics[width=0.3\textwidth, angle=0] {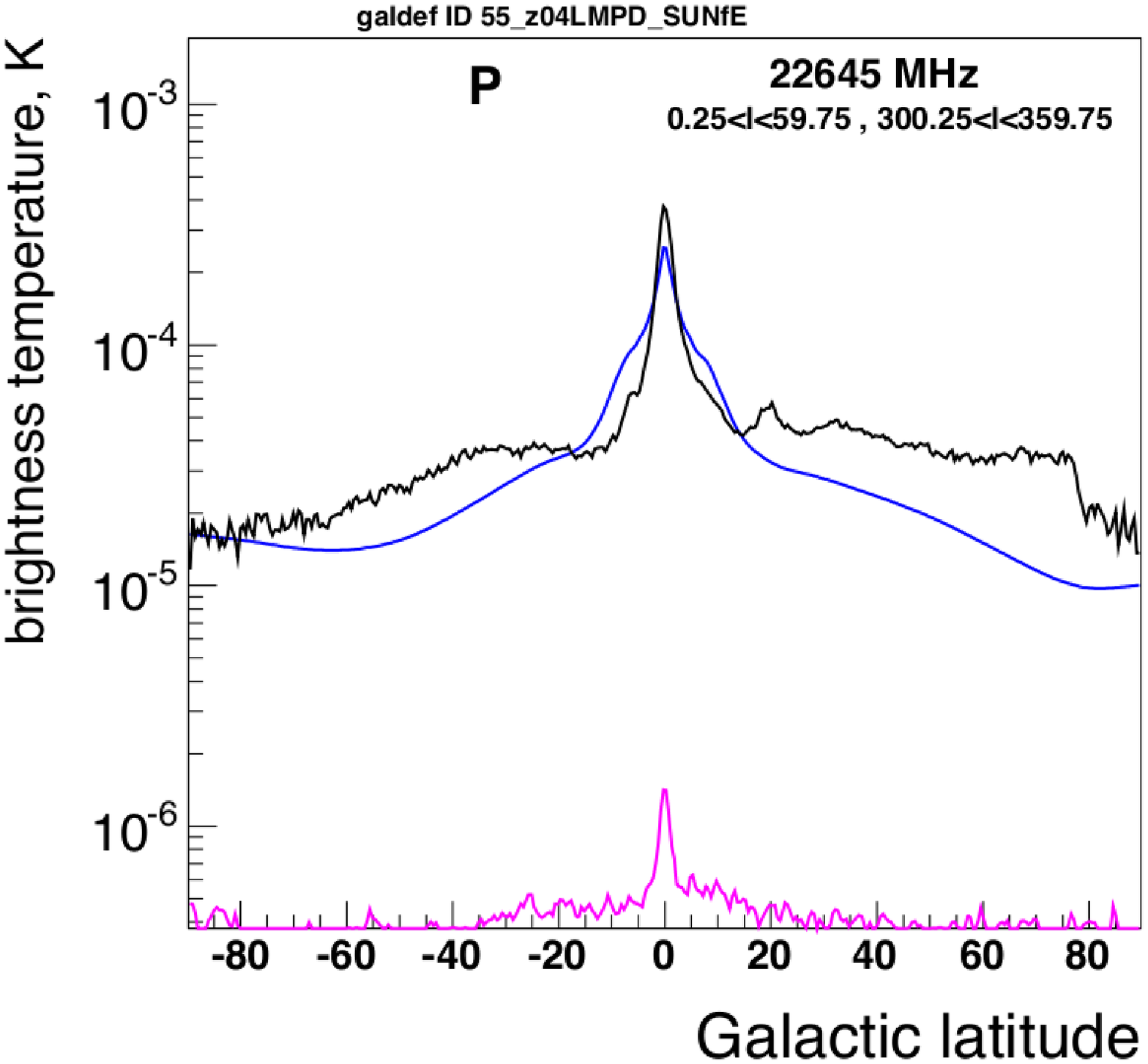}
\includegraphics[width=0.3\textwidth, angle=0] {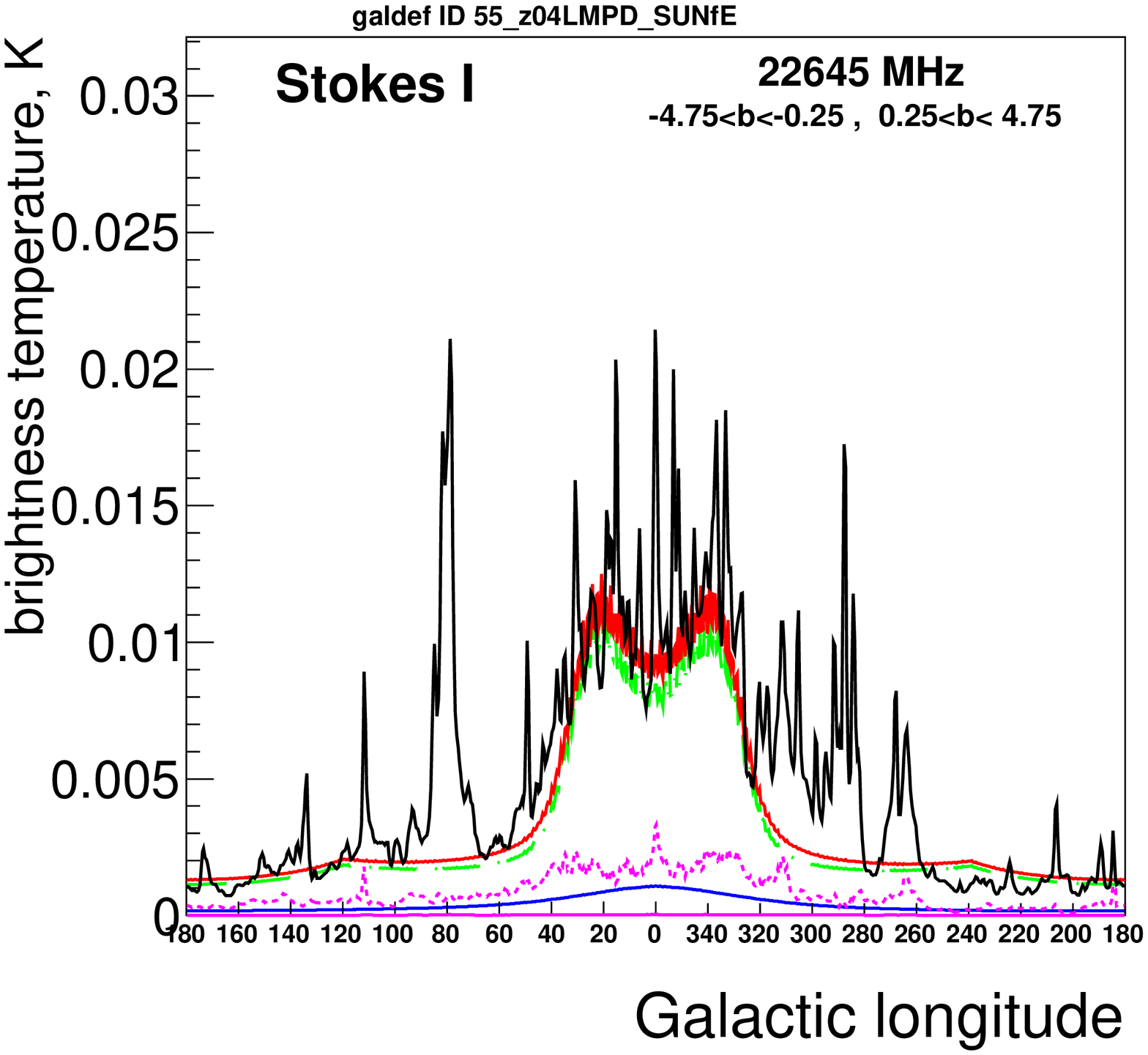}
\includegraphics[width=0.3\textwidth, angle=0] {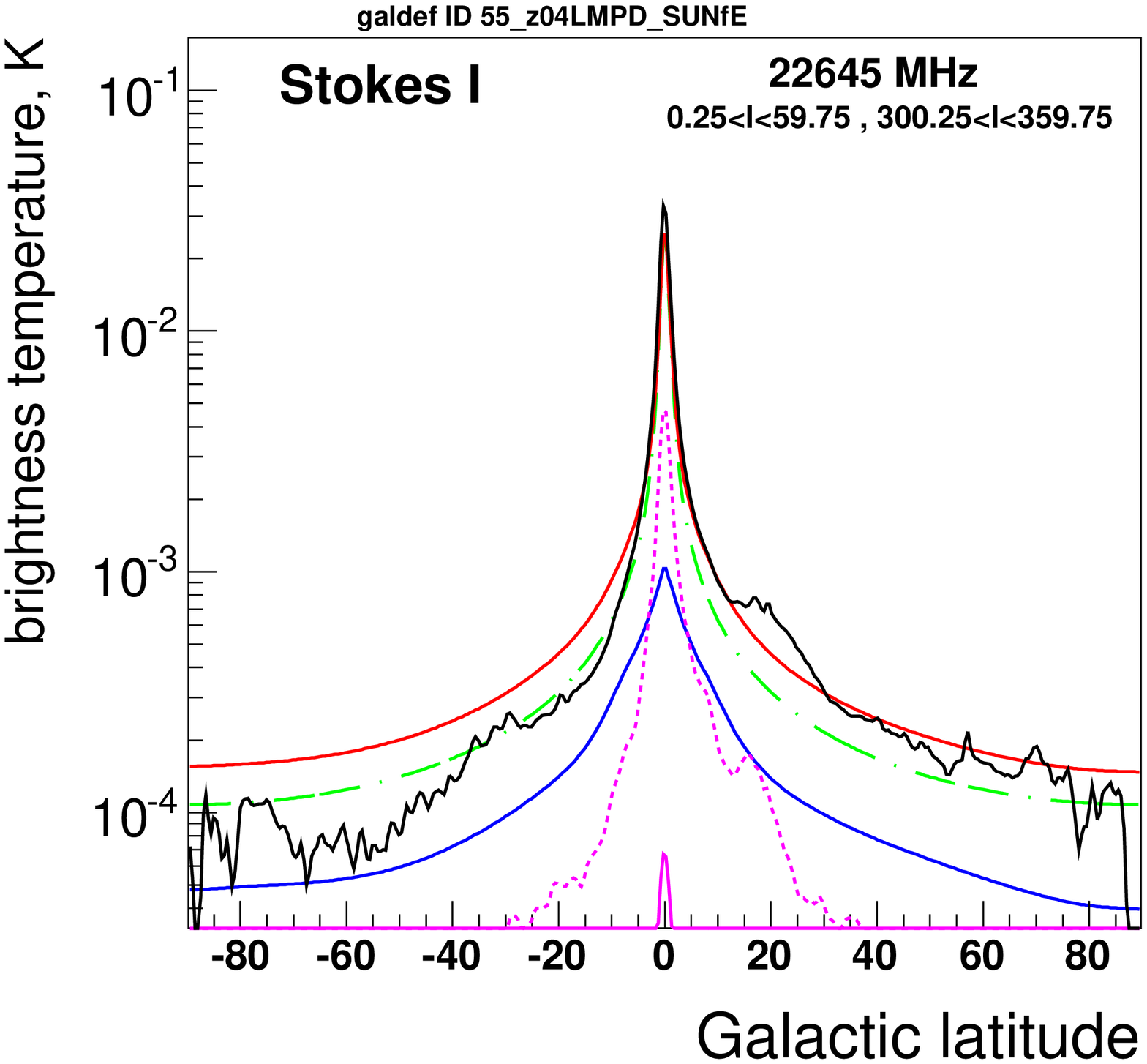}
\includegraphics[width=0.3\textwidth, angle=0] {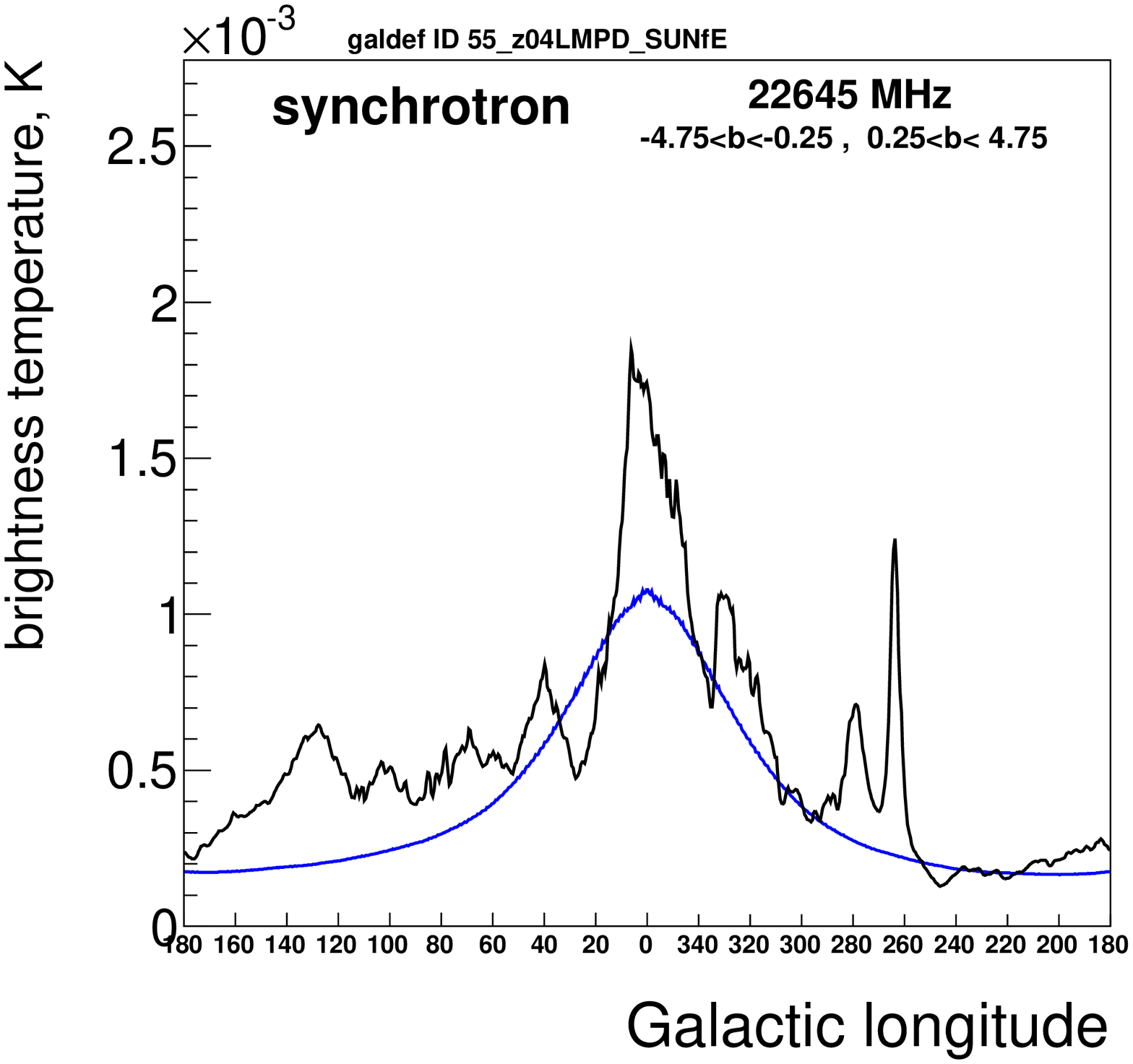}
\includegraphics[width=0.3\textwidth, angle=0] {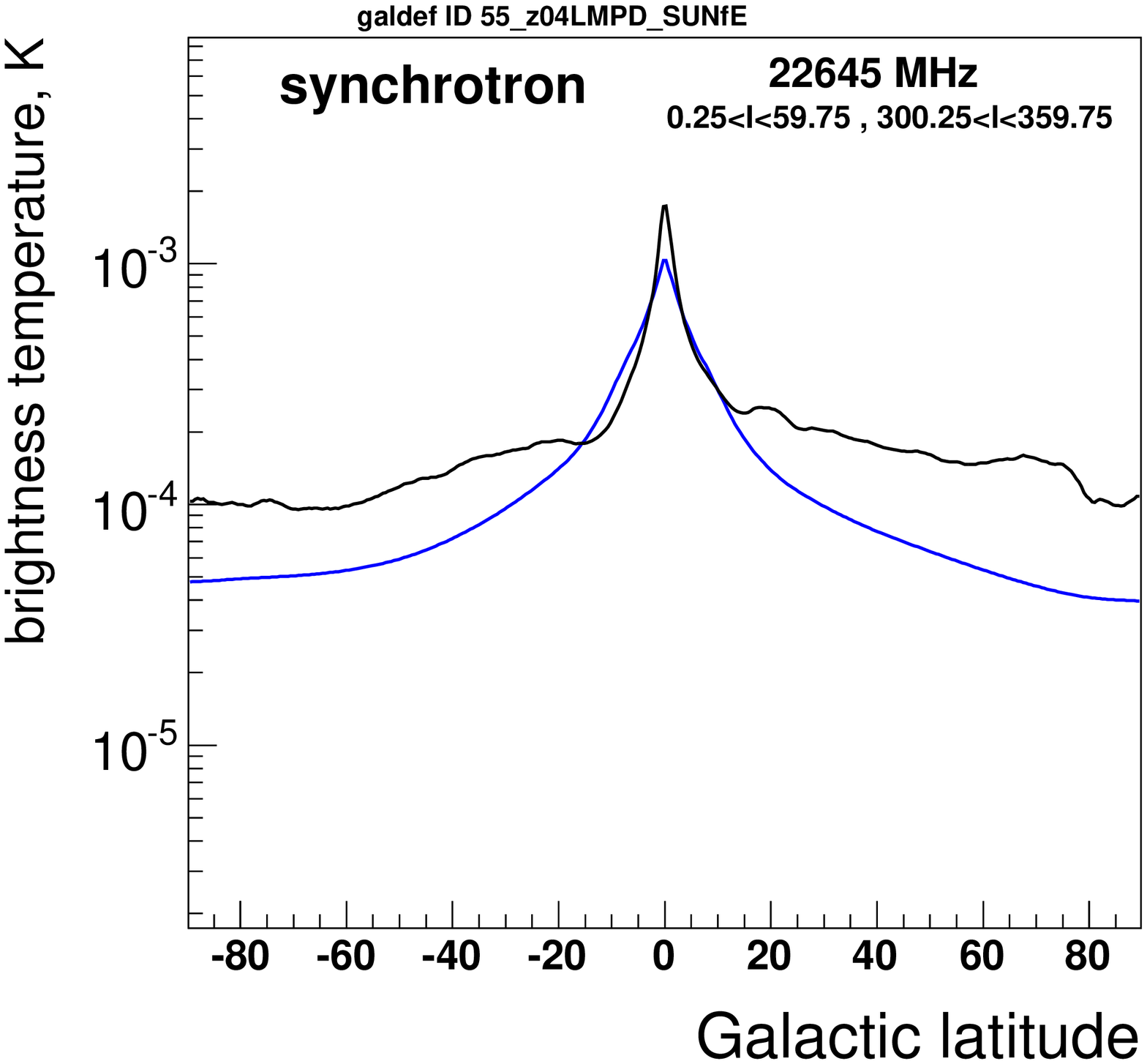}
\includegraphics[width=0.3\textwidth, angle=0] {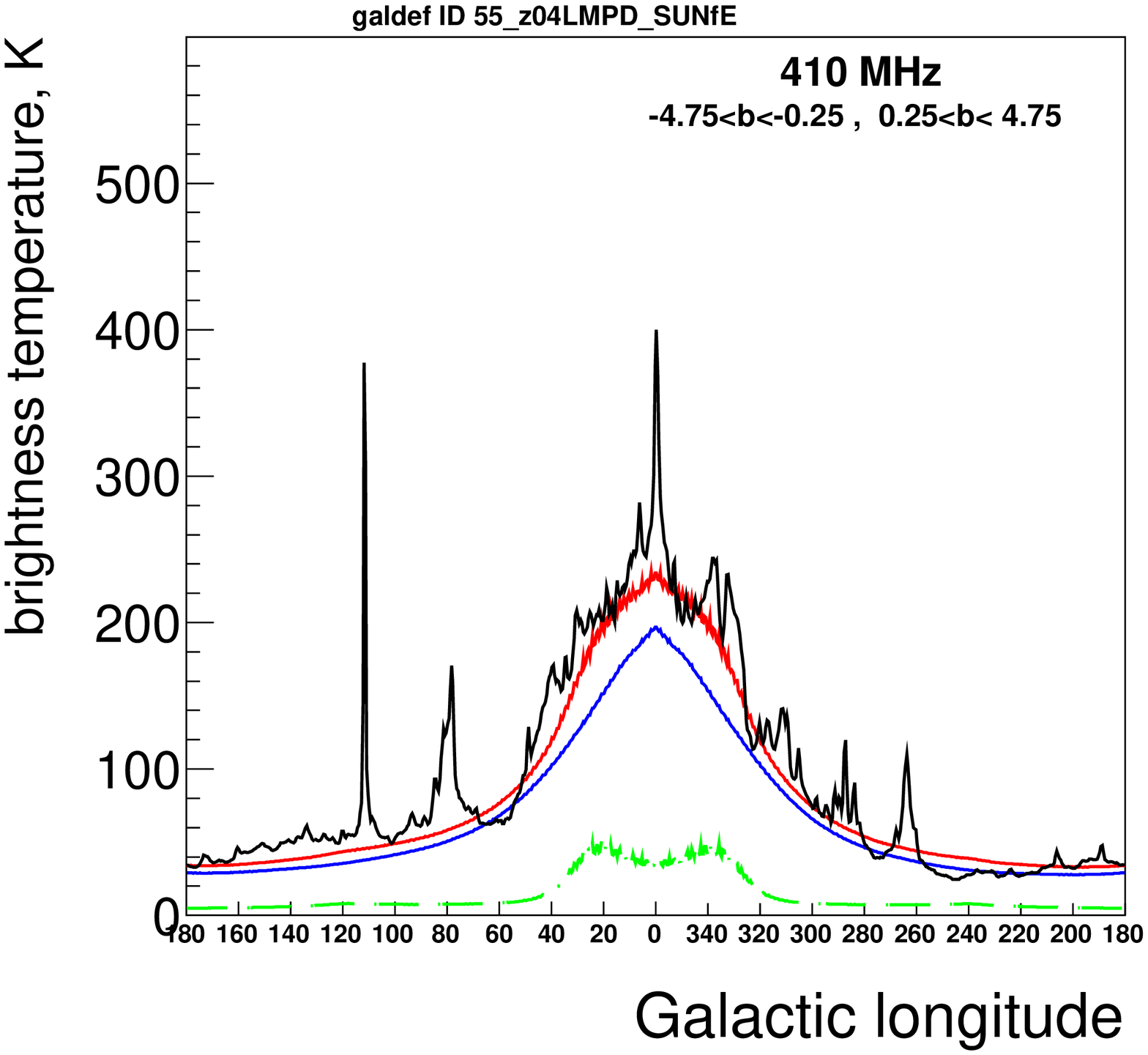}
\includegraphics[width=0.3\textwidth, angle=0] {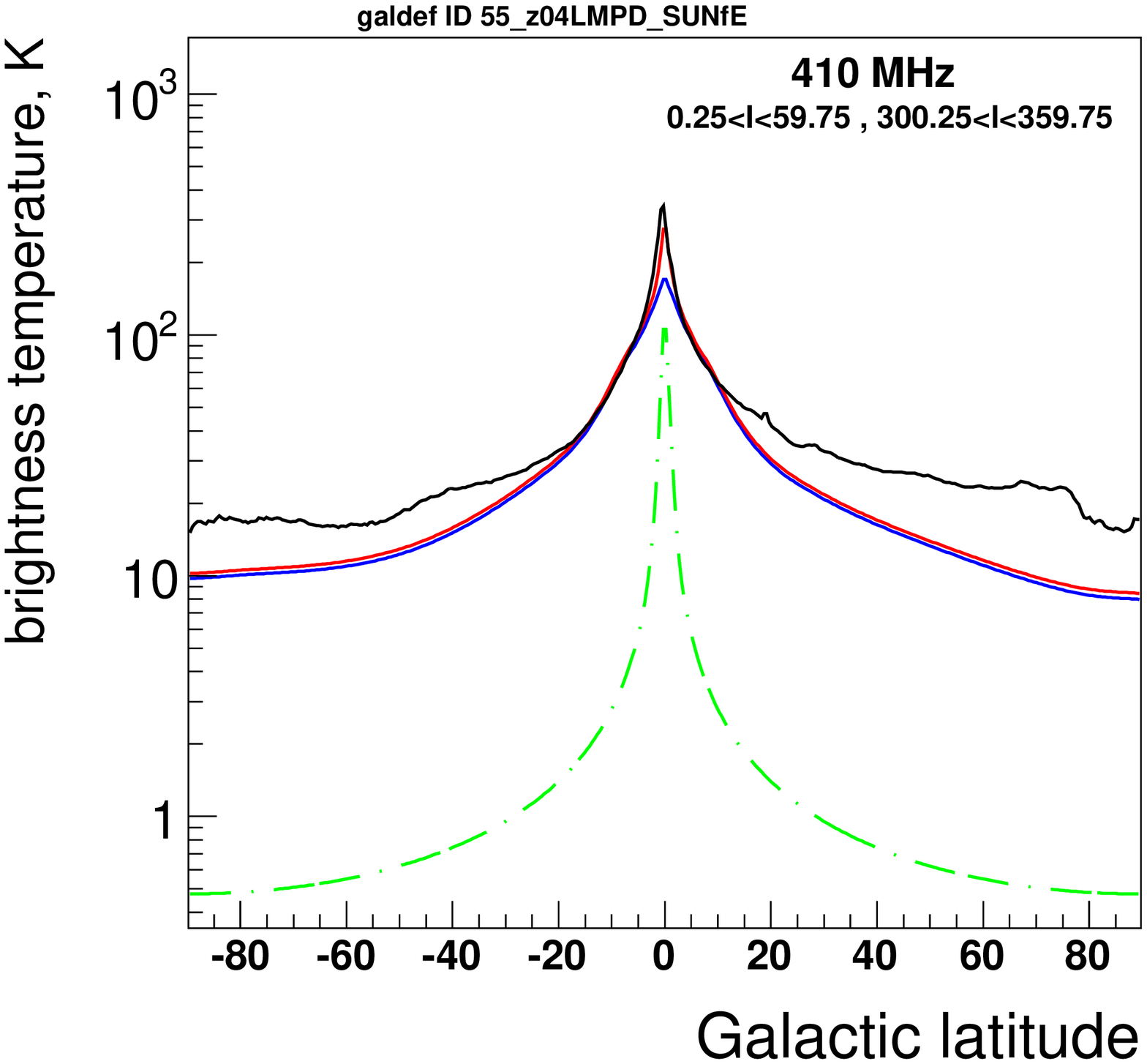}
\caption{ Model SUNE: intensity longitude and latitude profiles for B-field Model 1. The plots show the different model components: synchrotron (blue line), dust and spinning dust (pink dotted line), free-free (green dashed-dotted line), free-free plus synchrotron (cyan dashed line) and total (red line). Data are in black.  Left to right and top to bottom are longitude and latitude profiles for $Q$, $P$ and $I$ at 23 GHz, synchrotron-only at 23 GHz and 408 MHz. Longitude profiles are averaged over a region of $|b| \leq 5^\circ$, while latitude profiles over a region of $-60^\circ \leq l \leq 60^\circ$ around the Galactic Centre.  }
\label{SUN_prof}
\end{figure*}

\begin{figure*}
\centering
\includegraphics[width=0.3\textwidth, angle=0] {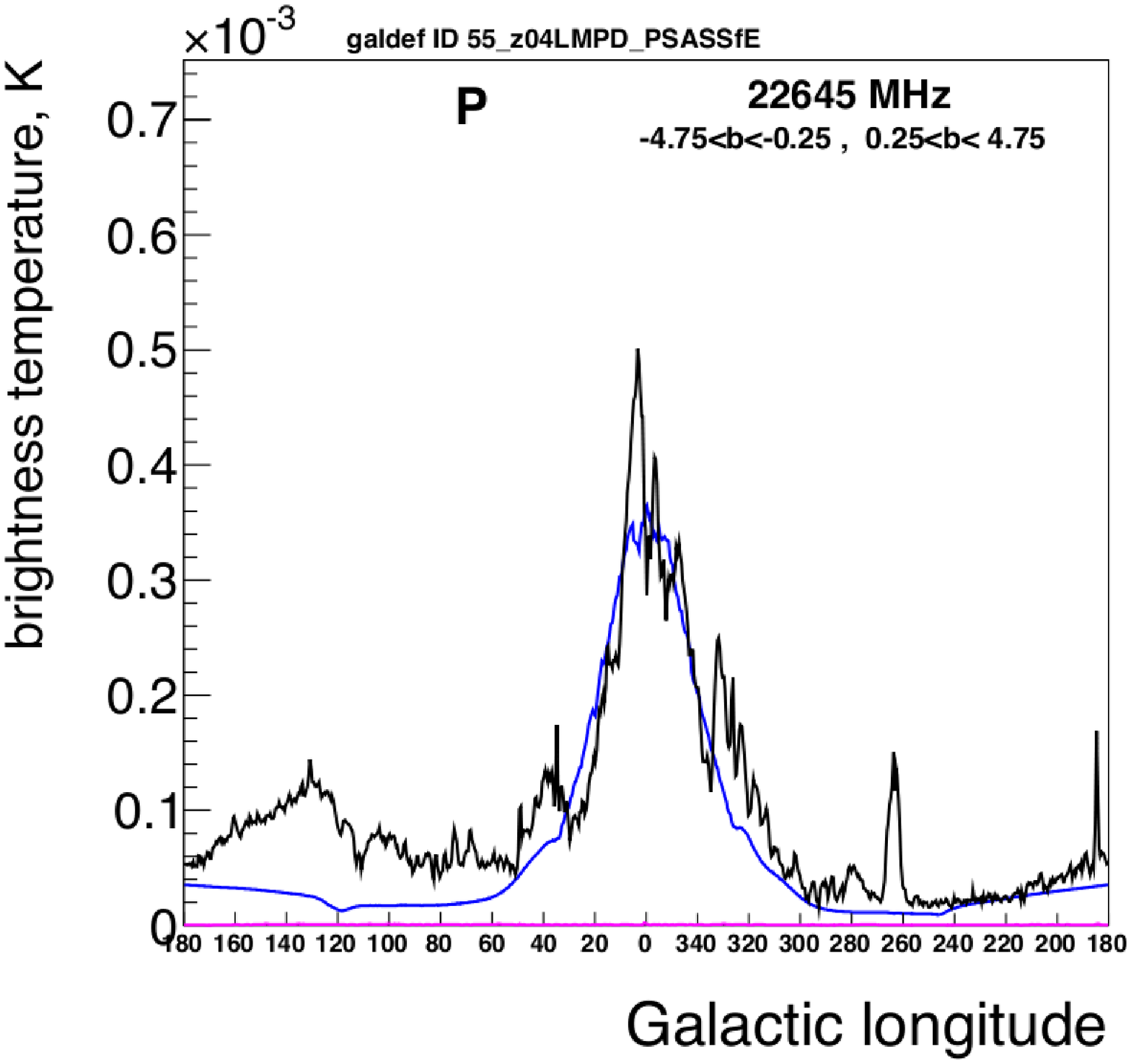}
\includegraphics[width=0.3\textwidth, angle=0] {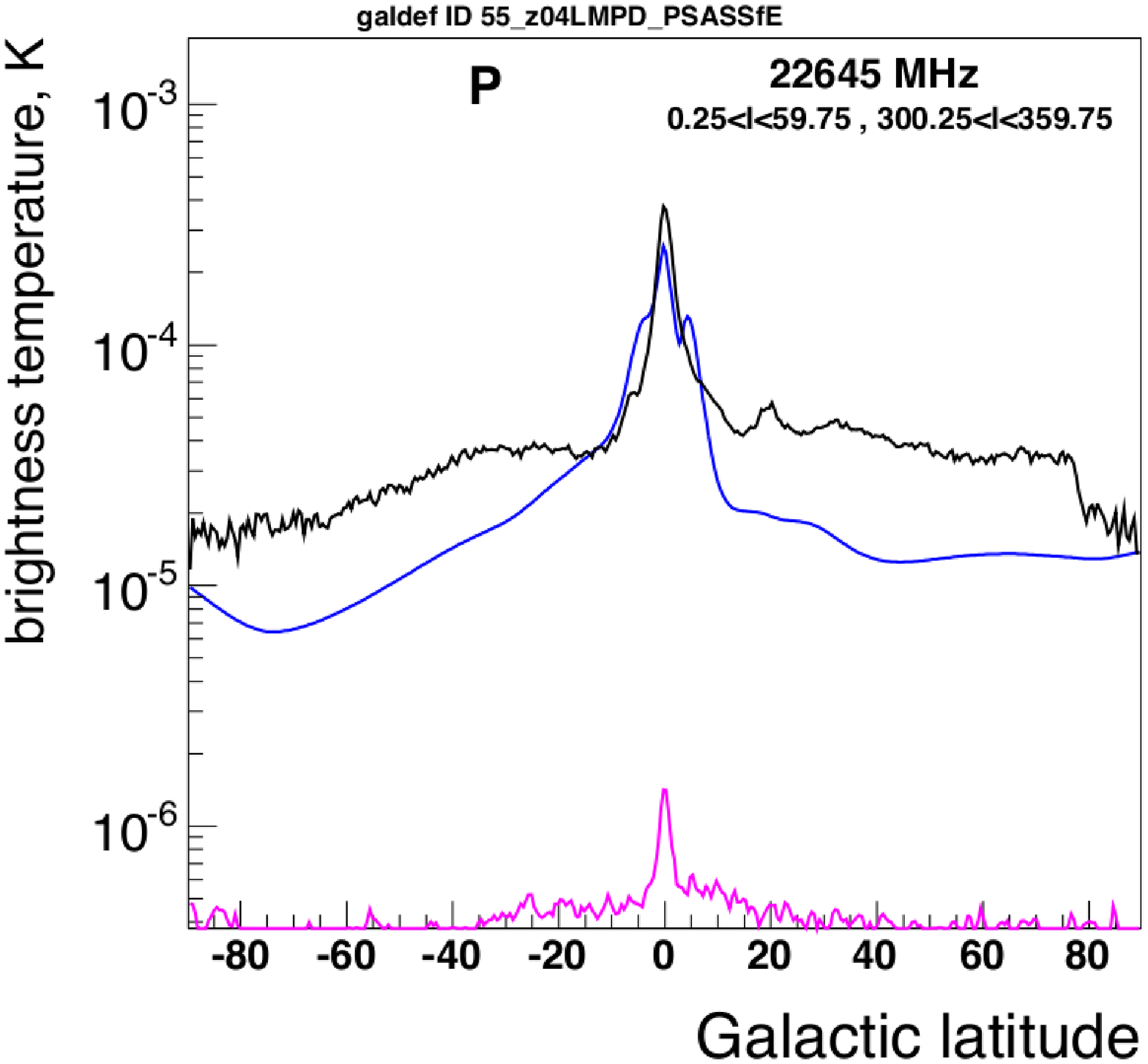}\\
\includegraphics[width=0.3\textwidth, angle=0] {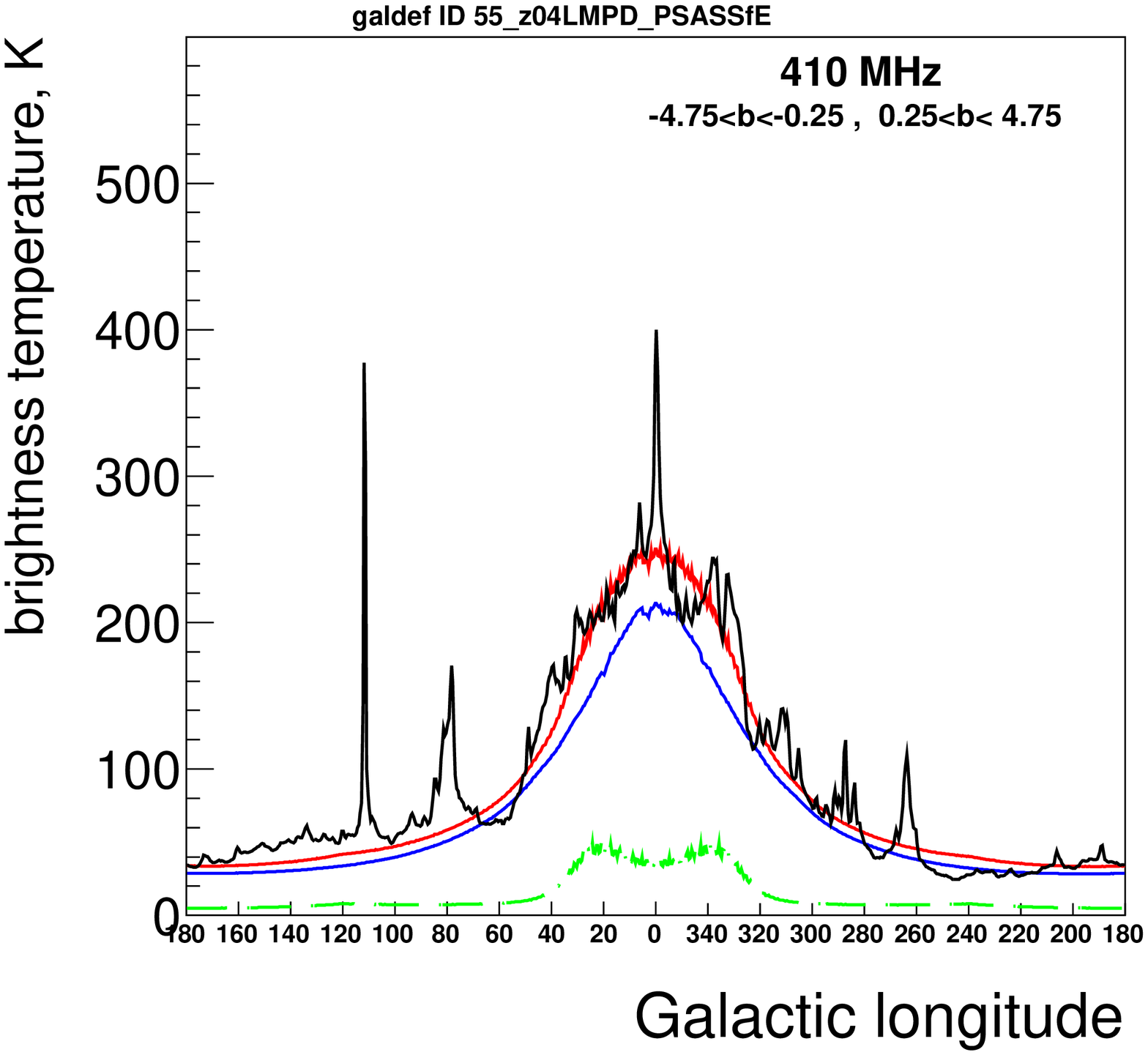}
\includegraphics[width=0.3\textwidth, angle=0] {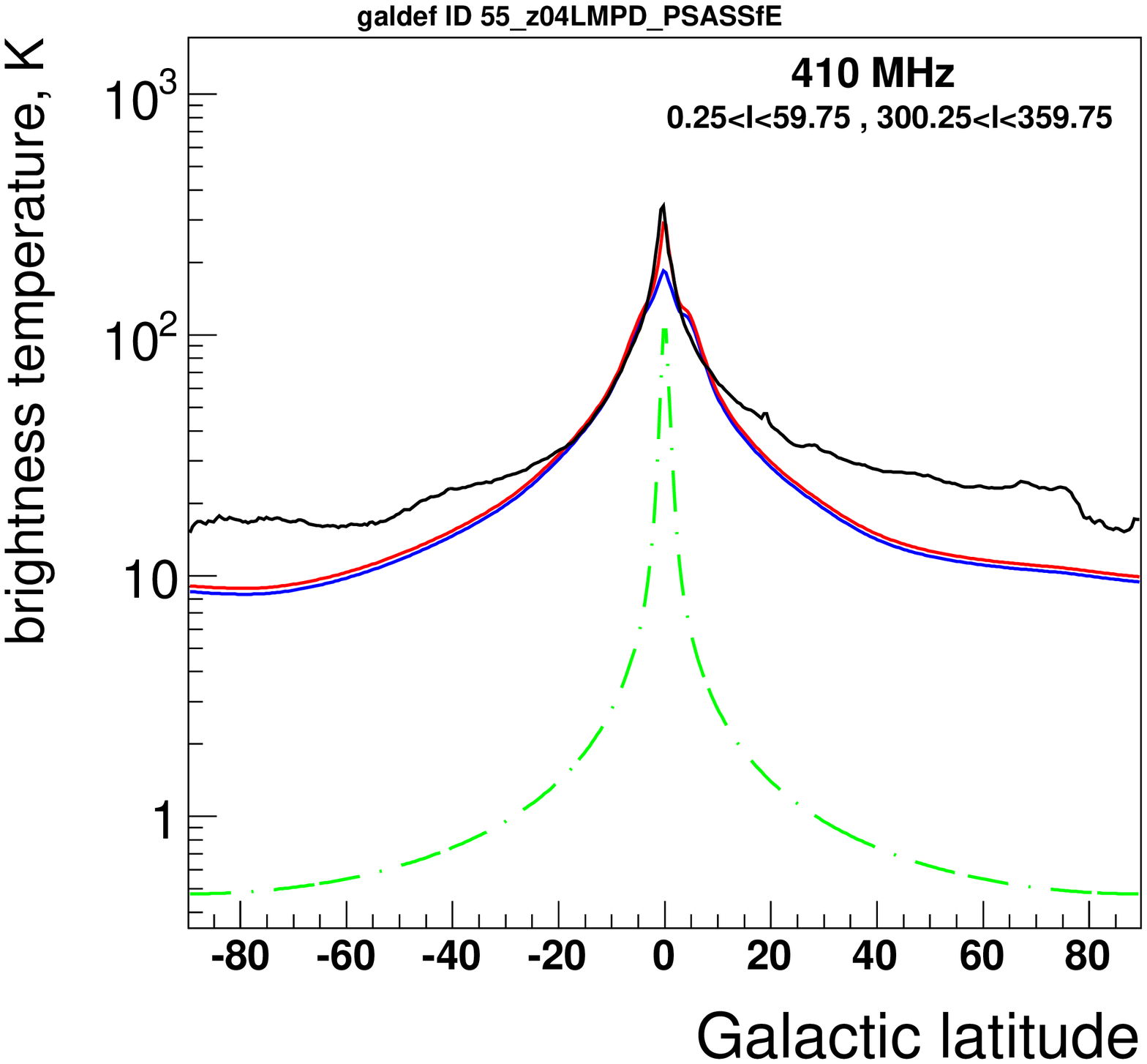}
\caption{  Model PSASSE:  brightness temperature longitude and latitude profiles for B-field Model 2.  For explanation see caption to Fig \ref{SUN_prof}. }
\label{ASS_prof}
\end{figure*}

\begin{figure*}
\centering
\includegraphics[width=0.3\textwidth, angle=0] {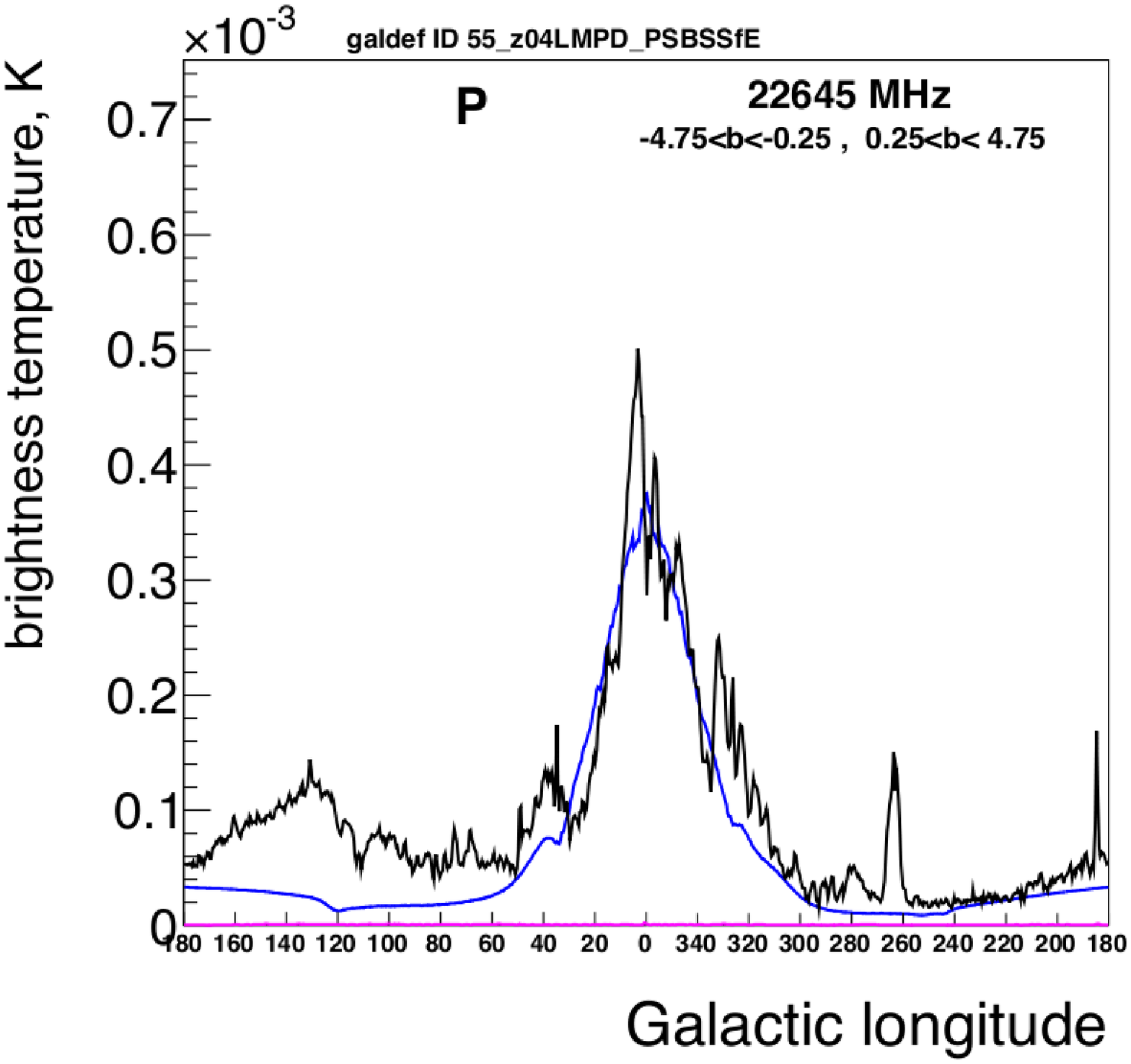}
\includegraphics[width=0.3\textwidth, angle=0] {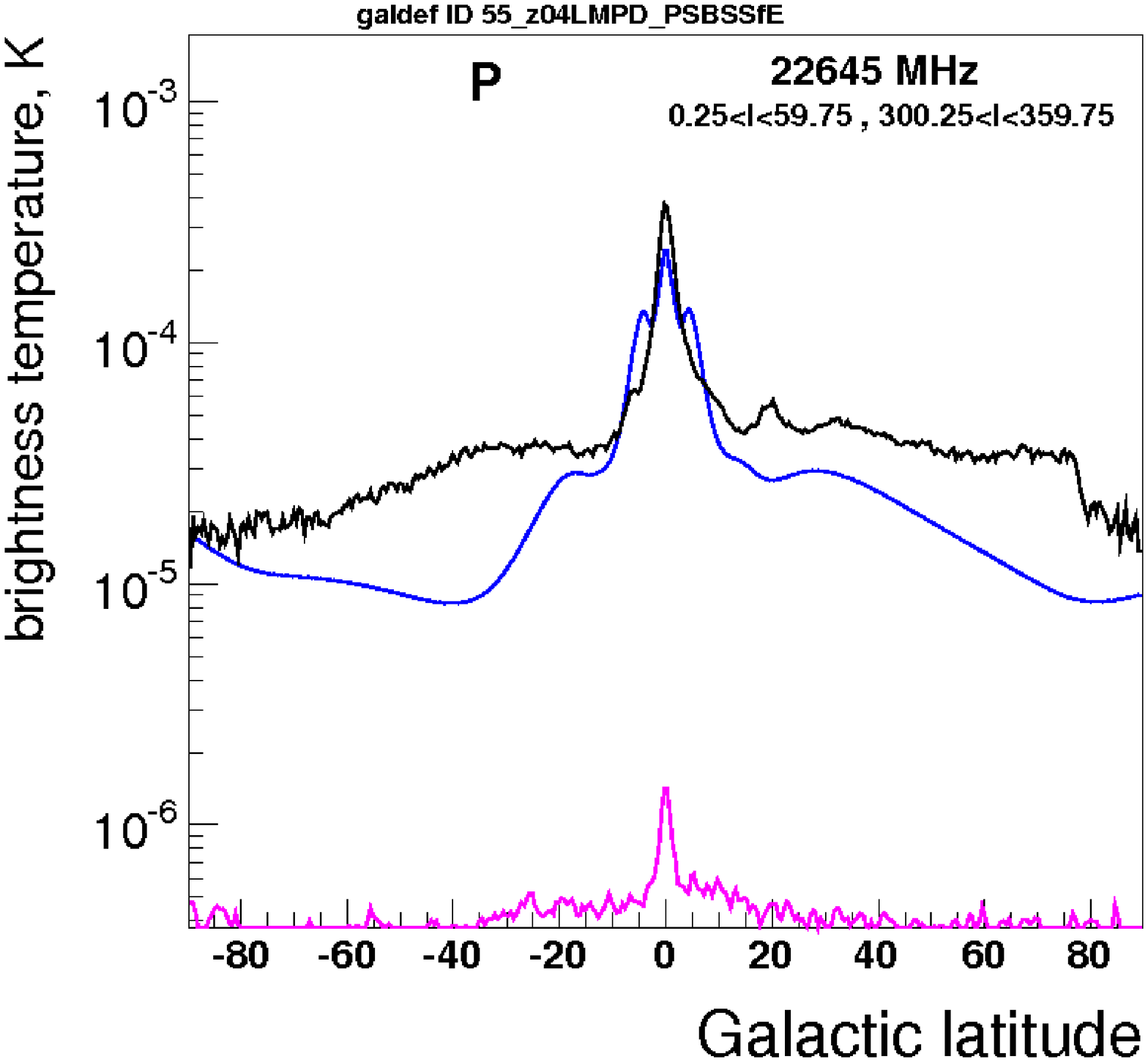}\\
\includegraphics[width=0.3\textwidth, angle=0] {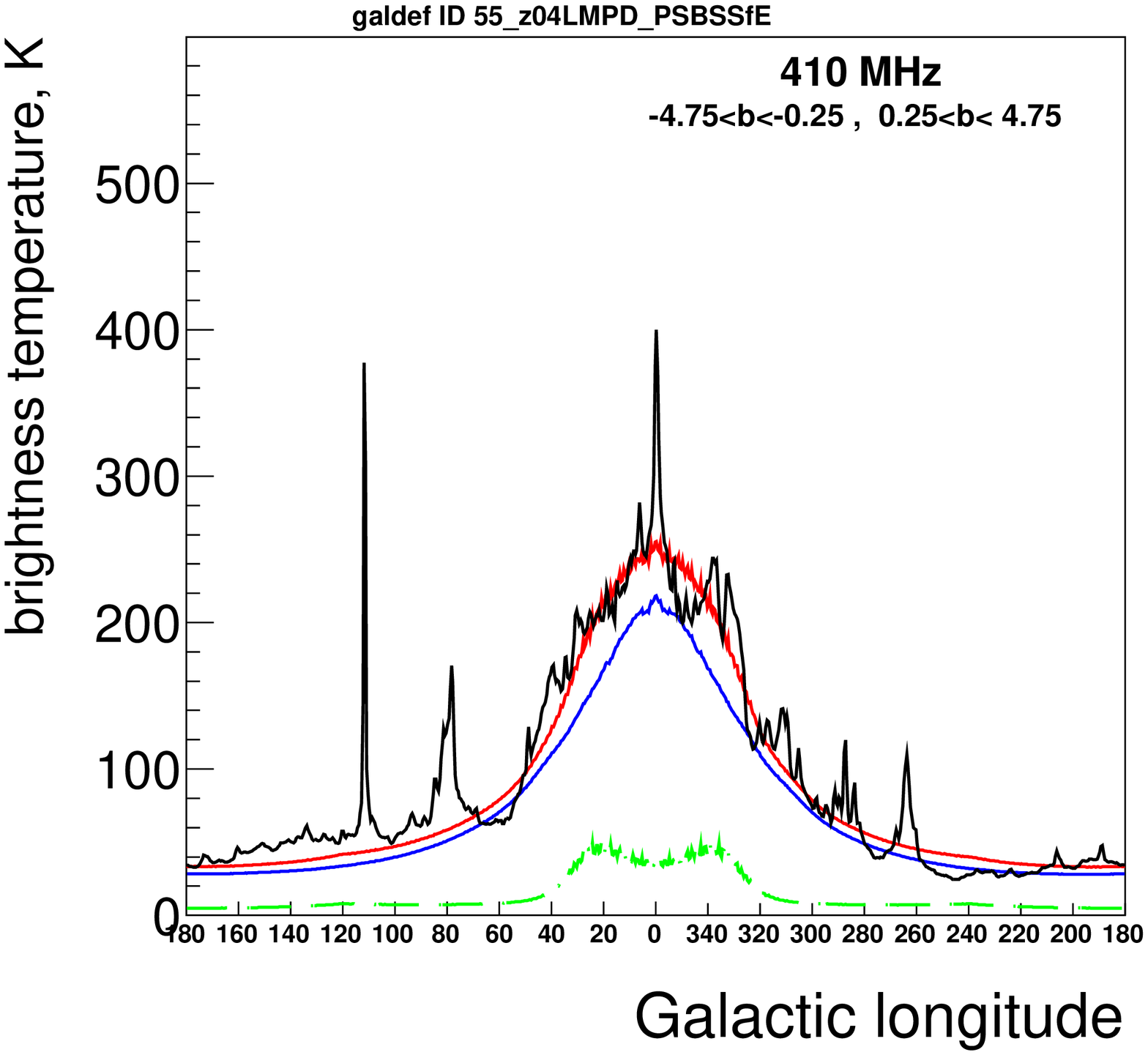}
\includegraphics[width=0.3\textwidth, angle=0] {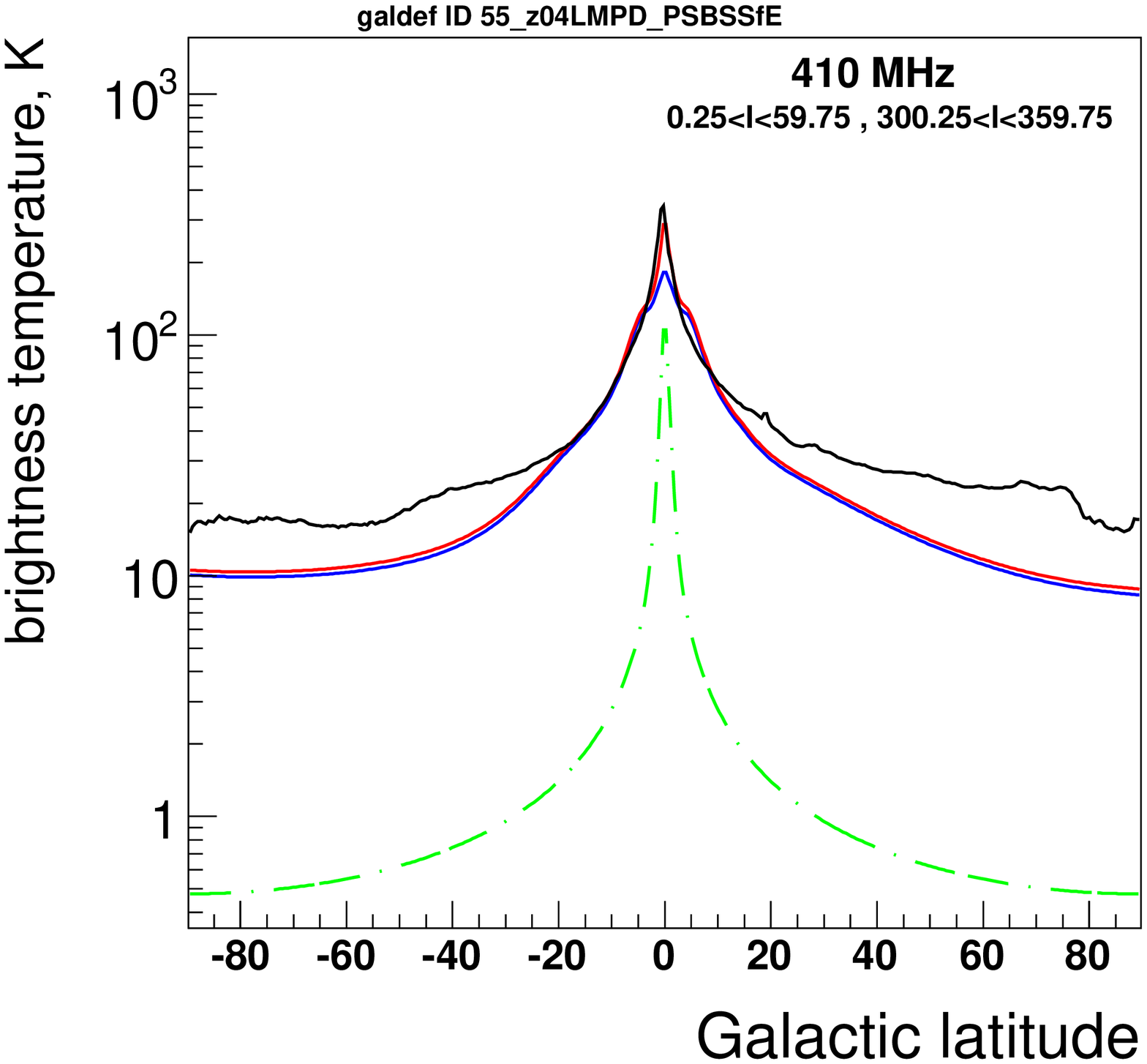}
\caption{  Model PSBSSE: brightness temperature longitude and latitude profiles for B-field Model 3. For explanation see caption to Fig \ref{SUN_prof}. }
\label{BSS_prof}
\end{figure*}

Figures \ref{SUN_prof}, \ref{ASS_prof}, \ref{BSS_prof} show the best-fitting longitude and latitude profiles for all models listed in Table \ref{Table1}. 
 We show longitude profiles in the plane for a region within $\pm 5^o$ latitude. 
 Latitude profiles are shown for a region of $60^o$ around the Galactic Center. 
Note that in the plots no further offsets, as the one reported in Table \ref{Table1}, have been included.
 In this way, profiles  can be directly compared to see the effect of different B-field models.

We find that:
\begin{enumerate}
\item An anisotropic random B-field has to be included in order to fit the polarization data.
\item The intensity of the anisotropic component is a factor 0.7 -1.1 the original intensity of regular B-field, depending on models.
\item The random  B-field with 4.7 - 5.3 $\mu$G reproduces  the data best.
This value is lower than the one found in SOJ2011 (7.5$\mu$G), where anisotropic random component was not accounted for.
\item The inner Galaxy peak and the longitude profiles in the plane are in general well reproduced. 
There is not much difference among the three different models in terms of large-scale description. 
This means that both random and anisotropic random components are reasonable modelled in the disc.  
However, Model 1 seems to reproduce better the overall data. 
In Models 2 and 3 in the $P$ longitude profiles, a double-sided peak is present due to the different regular field topology with respect to Model 1. 
In general P is overpredicted at intermediate latitudes when assuming the anisotropic random component with the same distribution of the regular one in the halo.
\item The latitude profiles of {\it WMAP} synchrotron, 408~MHz and $P$ show that we are underestimating the data for high latitudes for all models.
\item Galactic latitude profiles of $I$  show that the free-free model is not a good representation of the data at high latitude. 
However, it is not the purpose of this paper to attempt to improve this aspect, and this does not affect our conclusions.
\end{enumerate}

Given the freedom to add an anisotropic random field component, there is an apparent preference (judged by the relative $\chi^2$ values) for Model 1, in both 23 GHz $P$ and 408 MHz data (see Table \ref{Table1}).
This cannot be regarded as decisive, because all models have substantial deviations from the data in some regions.
However, we take this best model as reference for the following investigations.


\subsection{Effects of halo size} \label{halo}

The cosmic-ray halo size of our Galaxy is uncertain. 
Recent gamma-ray data seem to suggest a better fit to a model with a halo height  larger than 4 kpc \citep{diffuse2}.
We investigate the effects of increasing the halo size from 4 kpc to 10 kpc. 
A 10 kpc halo is still consistent with measurements of CR radioactive nuclei.

\begin{figure}
\centering
\includegraphics[width=0.3\textwidth, angle=0] {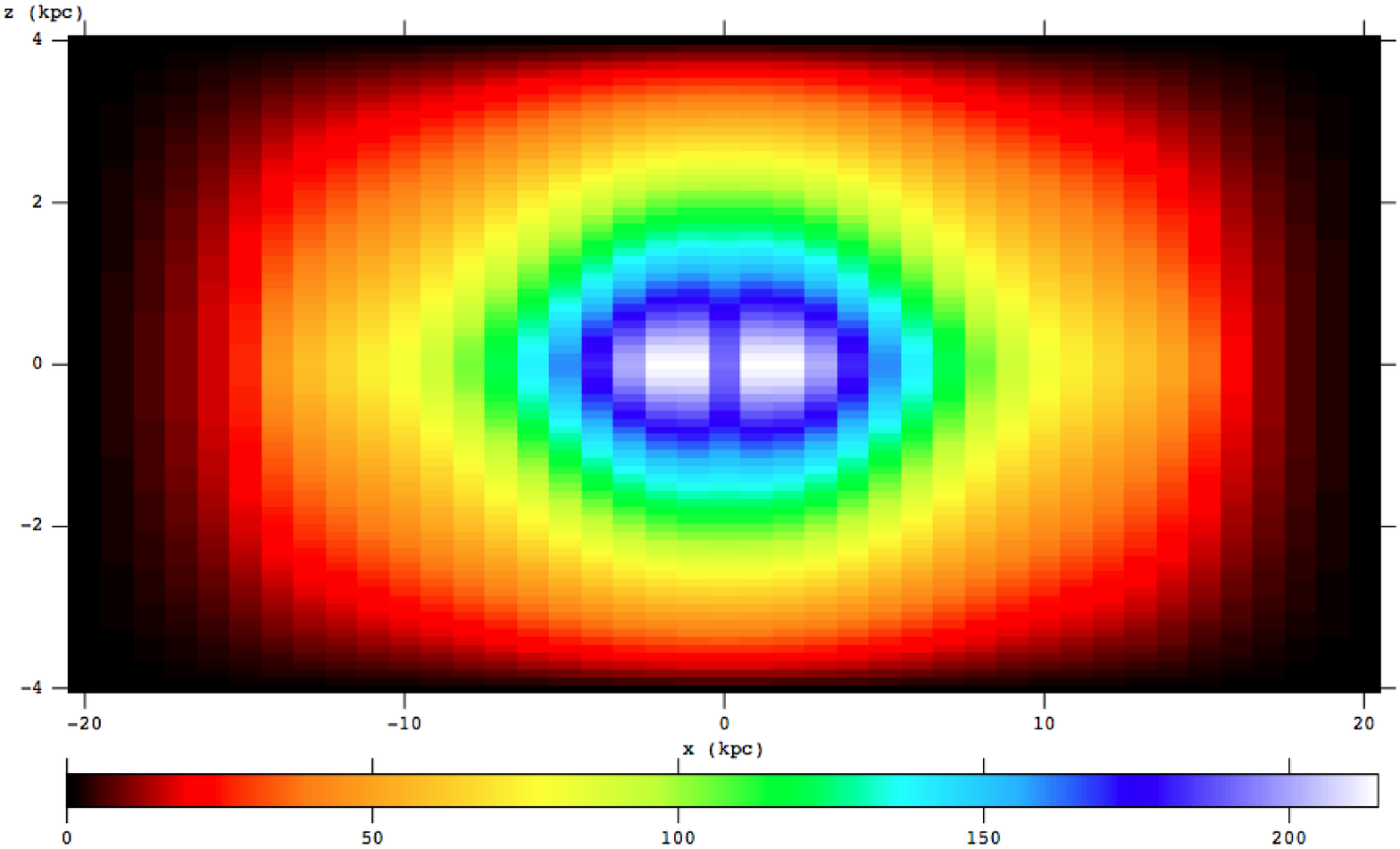}
\includegraphics[width=0.3\textwidth, angle=0] {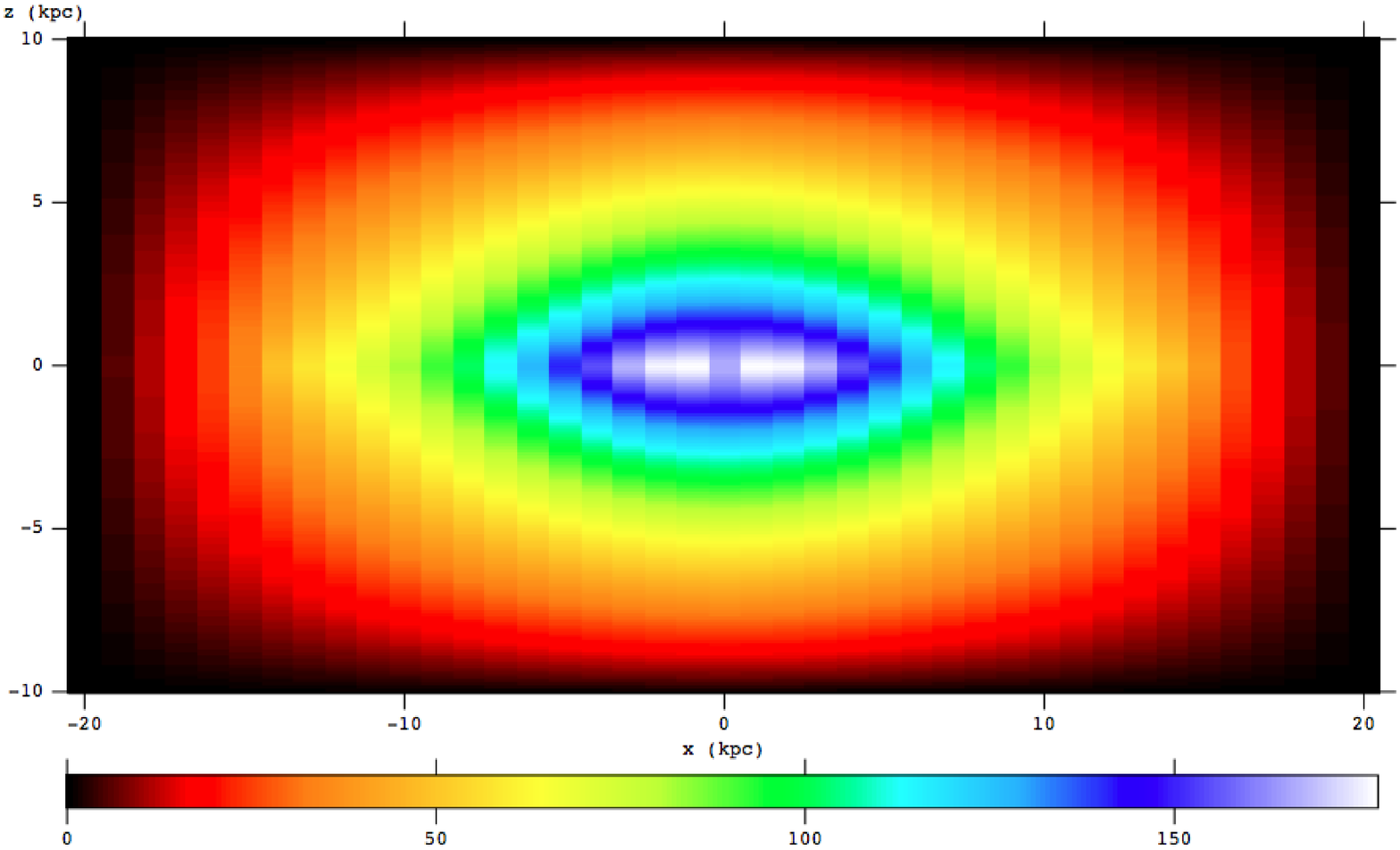}
\caption{Spatial distribution of electrons after propagation.
Section through the 3D model, in x and z directions for y=0.
 Energy is $\approx$ 1~GeV and density has arbitrary units.
Upper plot: halo height 4 kpc,
lower plot: halo height 10 kpc.
Note the differing $z$ scales.
}
\label{electrons_distribution}
\end{figure}

\begin{figure*}
\centering
\includegraphics[width=0.3\textwidth, angle=0] {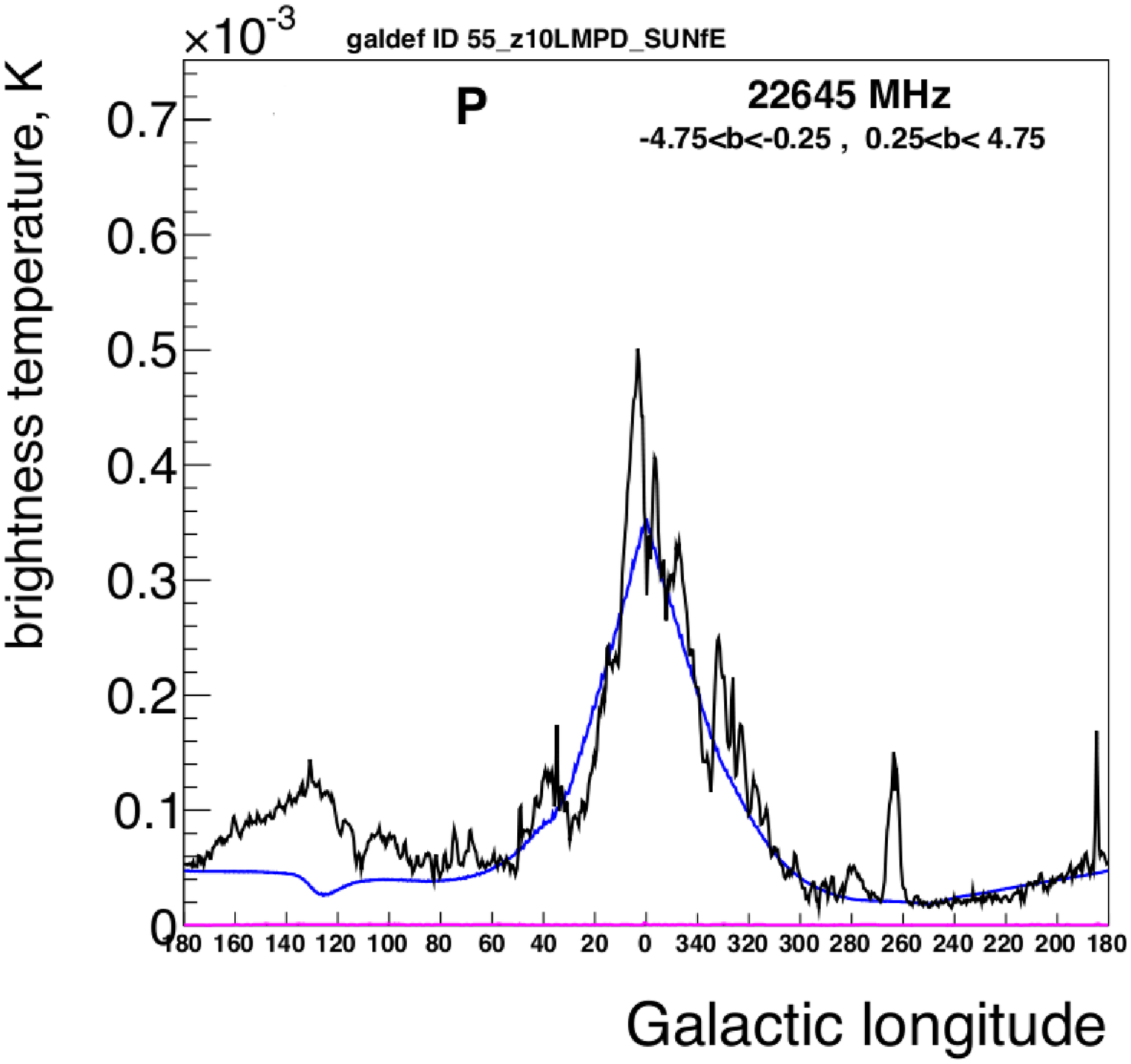}
\includegraphics[width=0.3\textwidth, angle=0] {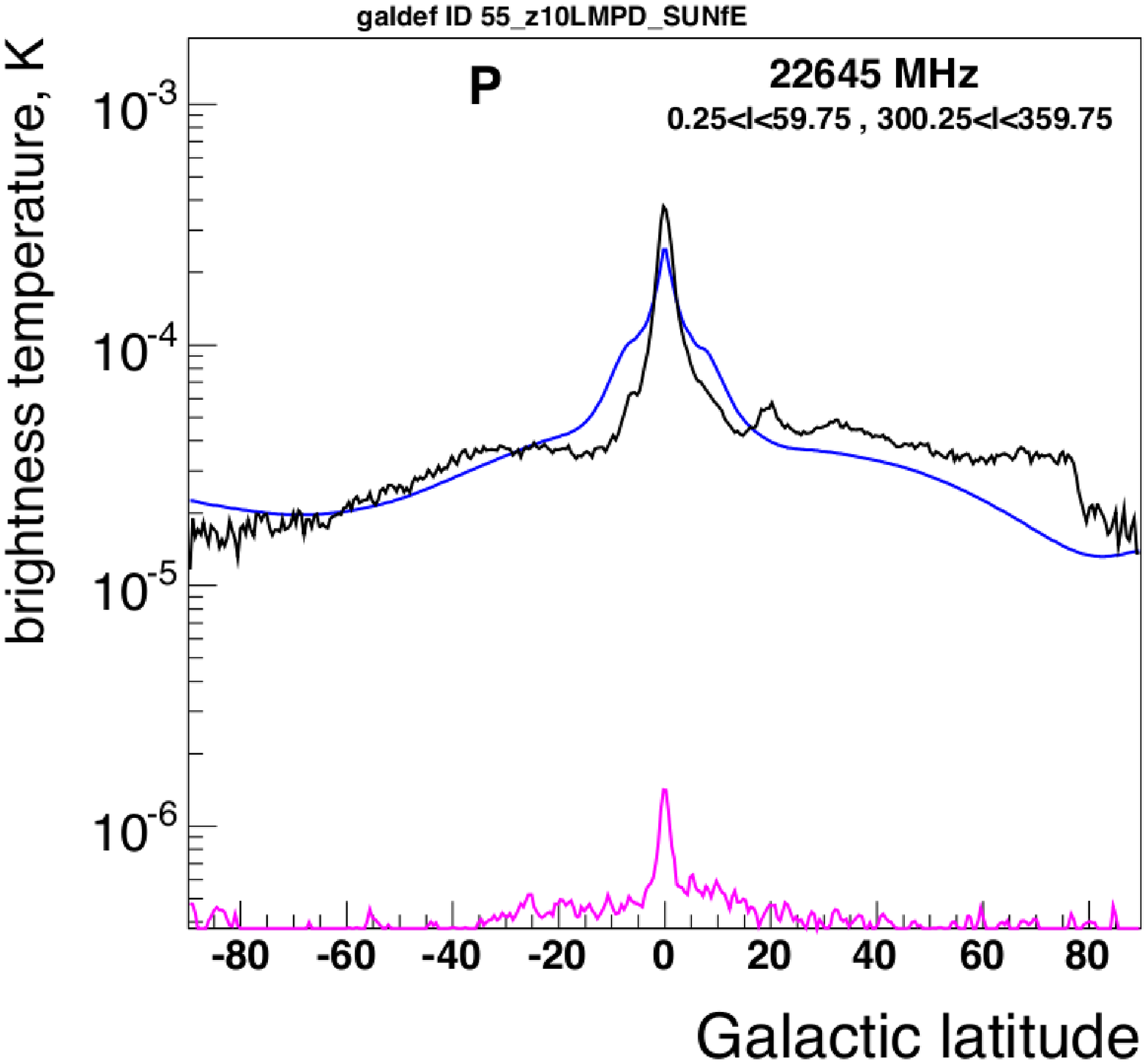}\\
\includegraphics[width=0.3\textwidth, angle=0] {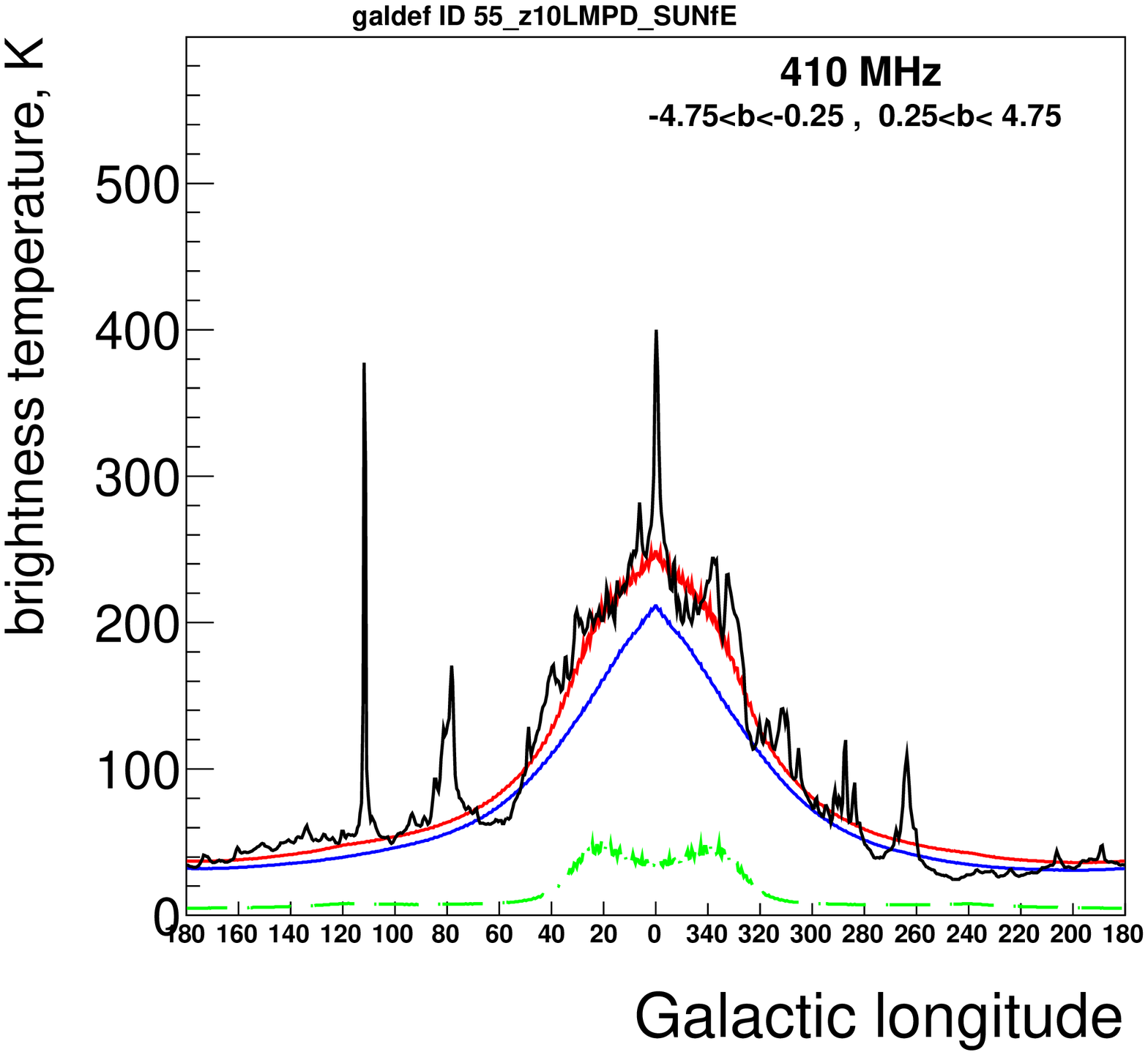}
\includegraphics[width=0.3\textwidth, angle=0] {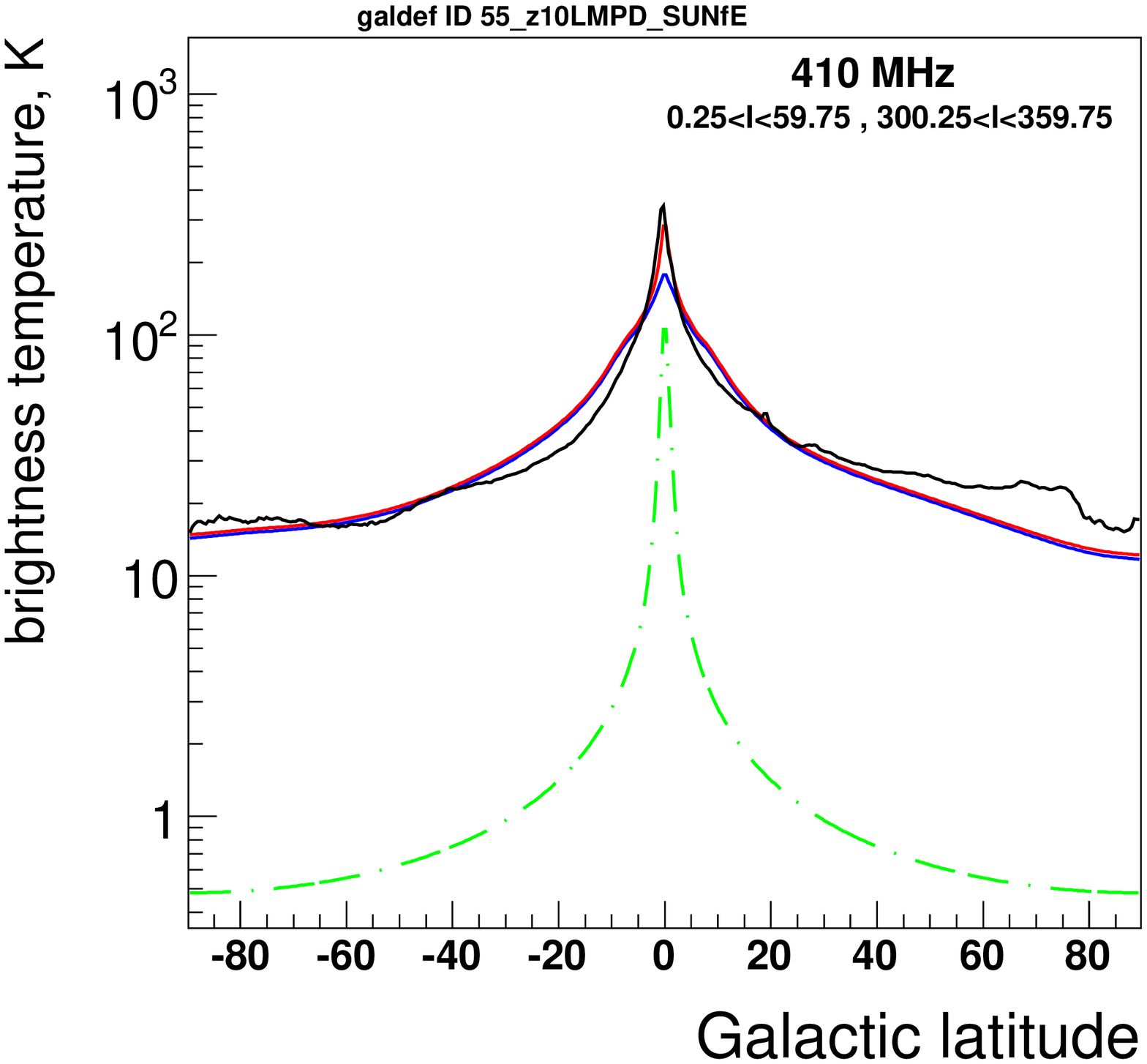}
\caption{ Model SUN10E:   brightness temperature longitude profiles for B-field Model 1 with  z = 10 kpc.  
 For explanation see caption to Fig \ref{SUN_prof}. 
}
\label{SUN10_prof}
\end{figure*}

In $SUN10E$ the halo size is taken as 10 kpc. 
Fig. \ref{electrons_distribution} shows the CR electron distribution in the Galaxy for
models $SUN04E$ and $SUN10E$ after propagation.
We first change only the halo height, maintaining the ordered (regular plus anisotropic random) B-field intensity fixed to the best-fitting values found for the $SUNE$ model (for 4 kpc).
 This allows us to evaluate the effects of increasing the halo size on the modelled profiles. 
Profile plots are shown in Fig.~\ref{SUN10_prof}, for comparison with Fig. \ref{SUN_prof} ($SUNE$). 
As expected, the emission in the plane does not change significantly, while a major effect is visible out of the plane with an associated increase also at intermediate latitudes, due to the longer integration along the line of sight.

In general, the systematic increase of the integrated synchrotron emission with latitude for larger halo sizes has the effect of flattening the latitude profiles above $20^o$,  which improves the high-latitude agreement with data. 
 This suggests that a halo size larger than 4 kpc is more likely.
  
We then perform the fit following the procedure described in Section \ref{fitting procedure}, leaving the B-field intensity free to vary. 
The best-fitting values are reported in Table \ref{Table2}.
We see that the larger CR halo (10 kpc vs 4 kpc) gives a worse fit (judged by the relative $\chi^2$ values between $SUNE$ and $SUN10E$).
In fact, for the case of $SUNE$, the best-fitting value of the offset is higher than for the case of $SUN10E$, suggesting that an isotropic component gives a better fit than a structured one for the whole sky.
The larger halo does give a better fit at very high latitudes, but it is too broad at intermediate latitudes, which leads to the worse $\chi^2$. 
This is caused by the fact the ordered (regular plus anisotropic random)  fields produce an excess at intermediate latitudes, which is more marked in Model 2 and 3, where secondary peaks arise at $~$10$^o$ latitude. 
This suggests that the toroidal halo field as adopted in the original models does not represent well the data at intermediate latitudes, if we assume that also the anisotropic random component has the same distribution.

Our best fitting model at 408 MHz, for the total synchrotron component, implies an offset between 1K and 8K, depending on the model. 
This is suggested by the $\chi^2$  (see Table \ref{Table1} and Table \ref{Table2})\footnote{Note that the offset was also allowed to assume negative values.}  
and it is visible in the latitude profile plots as a flat emission for $|b|>30^o$. 
A large offset indicates that the synchrotron emission is not well modelled. 
This is true for all our models except $SUN10E$. 
The same trend is seen in the polarization data. 
However, polarized data are affected by systematic bias at high latitudes,  which is not under our full control.
On the other hand  in the 408 MHz data, the zero-level uncertainty and the extragalactic contribution is reasonably well known\footnote{We correct the 408 MHz map for CMB, extragalactic non-thermal component and zero-level correction by subtracting 3.7K, as given by \cite{Reich1988}. This value could even be overestimated, since more recent works found 2.7K \cite{Reich2004} and 1.6K \cite{Guzman2011}.}. 
This again suggests the need for a CR halo height larger than 4 kpc in order to describe the high latitude emission.


\subsection{Effects of different CR  distributions}

\begin{figure}
\centering
\includegraphics[width=0.3\textwidth, angle=0] {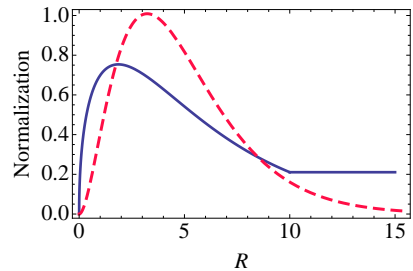}
\caption{CR source distributions from \citet{strong2010} (blue line) and pulsar-based \citet{lorimer} (red dashed line). 
 $R$ is the Galactocentric radius in kpc. The distributions are normalized at $R$ = 8.5 kpc. }
\label{CR_source}
\end{figure}

CR are believed to be accelerated by supernovae remnants (SNR). Their distribution is not well known. 
We  check  the influence on the synchrotron  modelling by varying the electron and positron distributions in the Galaxy. 
We do this by changing the distribution of the CR sources.  
We first consider CR sources as being distributed as measured for pulsars. 
This assumption is justified by the fact that pulsars trace SNR, which are believed to be the sources of CRs, and their distribution is better determined than SNR. 
We also use the distribution of CR sources given by \cite{lorimer} and used in \cite{diffuse2}. 
Both CR source distributions are shown in Fig.~\ref{CR_source}: that of \cite{strong2010} is flatter than the original \cite{lorimer}, but has more CR sources in the innermost region $R<2$~kpc. 

\begin{figure*}
\centering
\includegraphics[width=0.3\textwidth, angle=0] {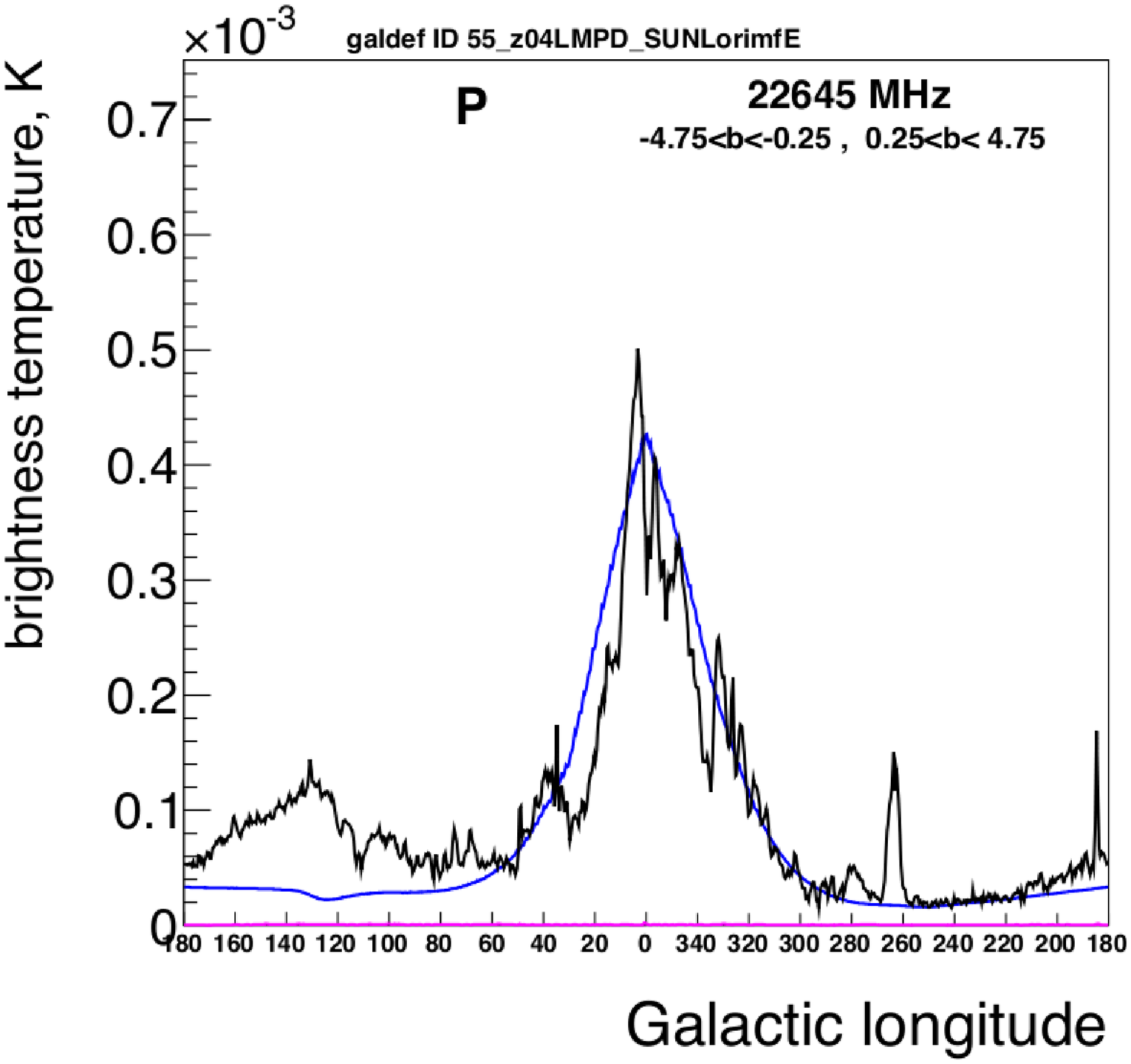}
\includegraphics[width=0.3\textwidth, angle=0] {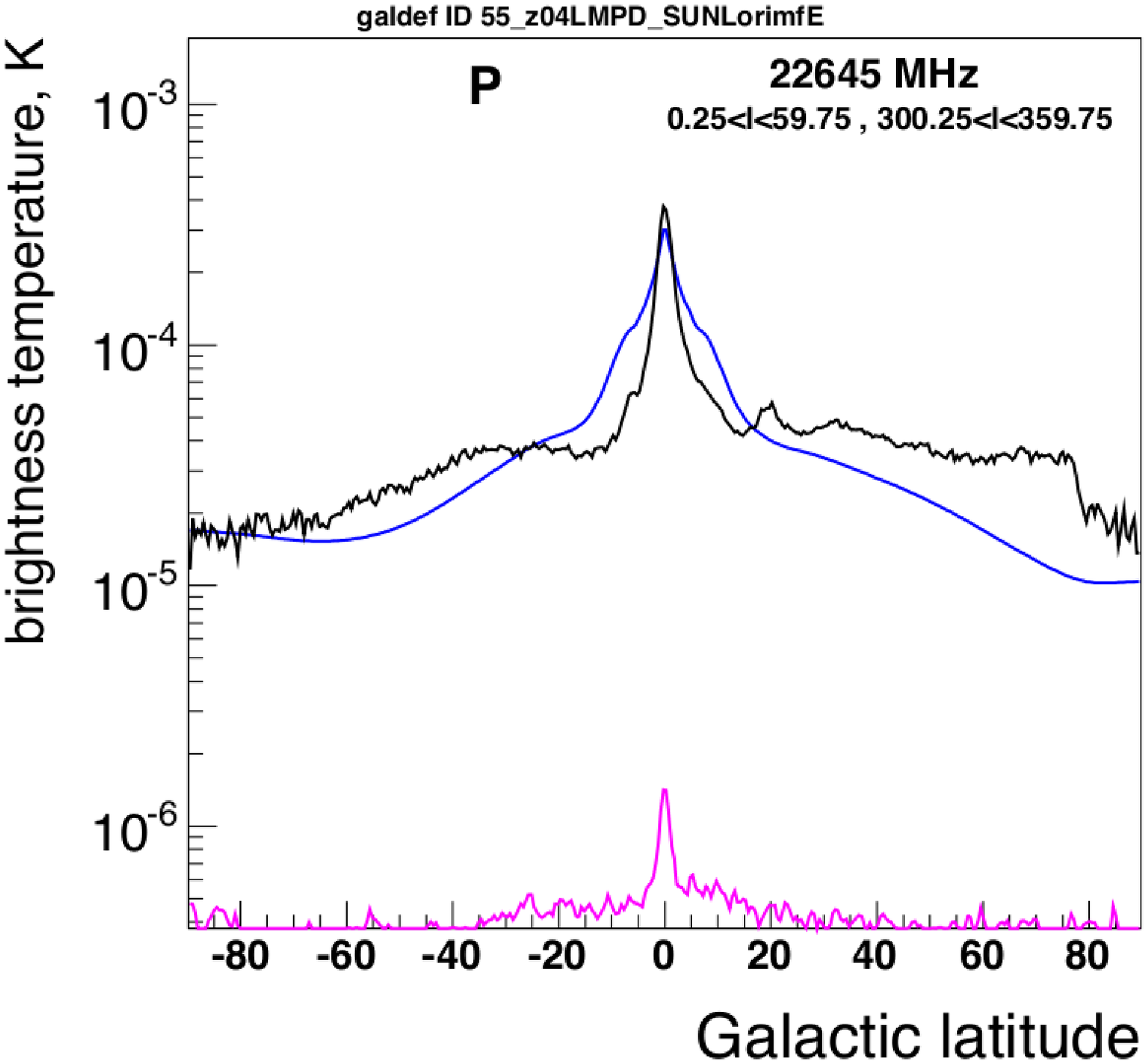}\\
\includegraphics[width=0.3\textwidth, angle=0] {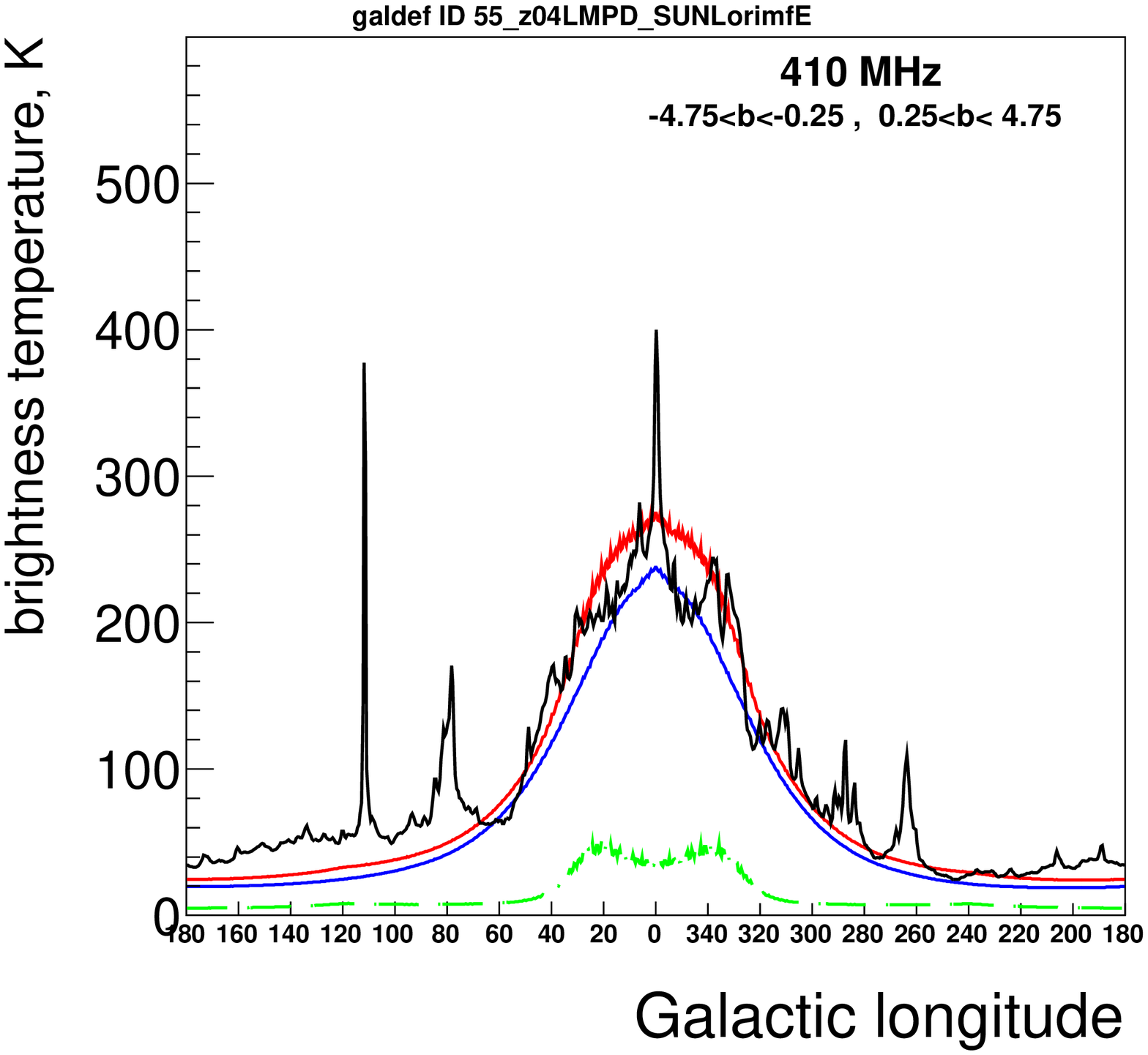}
\includegraphics[width=0.3\textwidth, angle=0] {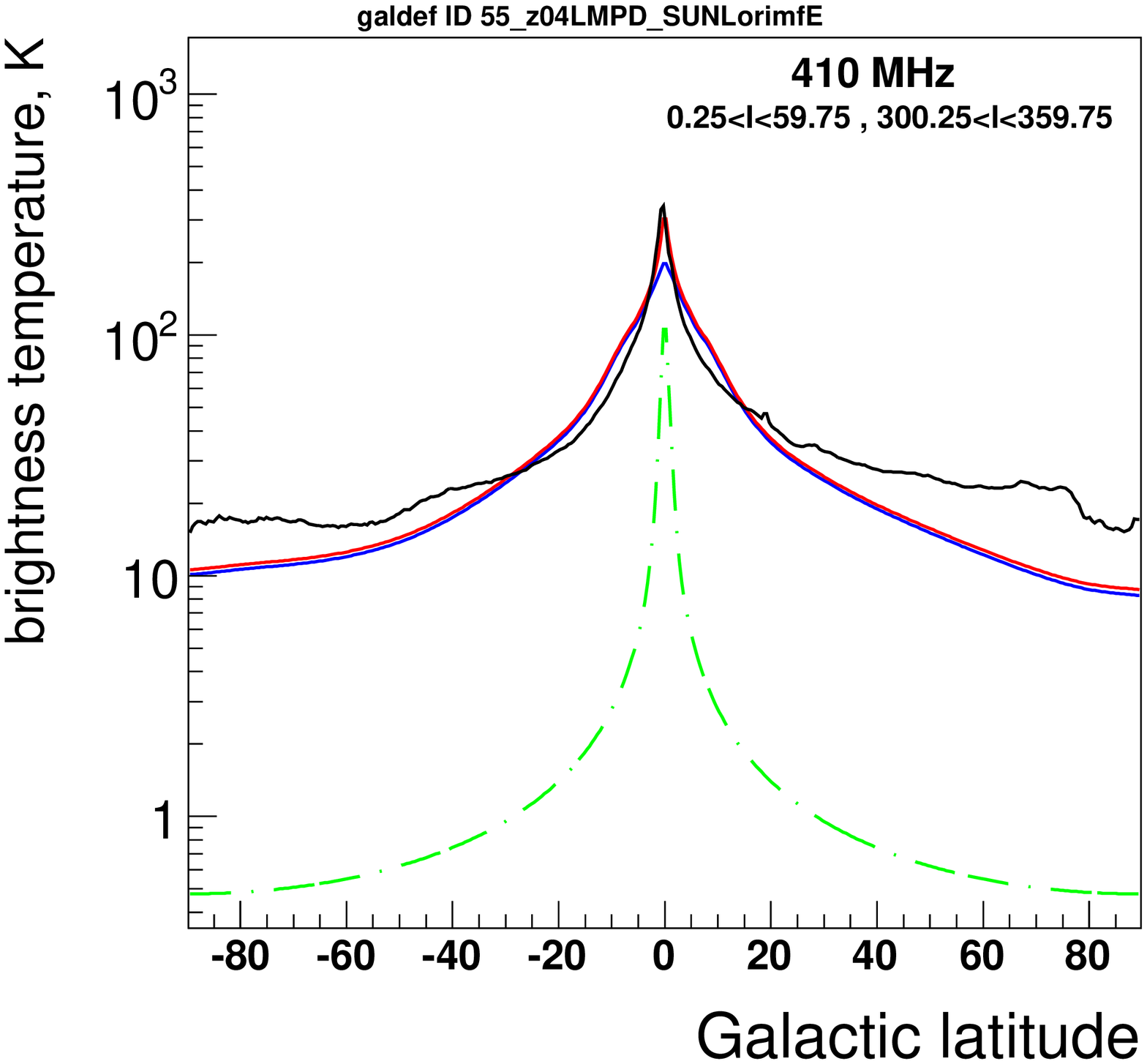}
\caption{Model SUNLorimE:    brightness temperature longitude and latitude profiles for B-field Model 1 and z=4 kpc. CR source distribution based on pulsars from \citet{lorimer}. For explanation see caption to Fig \ref{SUN_prof}. }
\label{SUNLorim_prof}
\end{figure*}

Varying the CR source distribution does not significantly affect other propagation parameters; hence, we maintain all the parameter values of $SUNE$.
The resulting model is called   $SUNLorimE$.
We use Model 1 for the B-field, but most effects apply to the other B-field models, given the same CR source distribution.
The corresponding  profiles are presented in Fig. \ref{SUNLorim_prof}, to be compared to Fig. \ref{SUN_prof}. 

The most relevant effect can be seen in the 408~MHz and 23~GHz synchrotron-only longitude profiles, while small changes with respect to Fig. \ref{SUN_prof} are seen in  the latitude profiles. 
 We see that the model intensity in the inner Galaxy region is $\approx$ 20 per cent higher than that in Fig. \ref{SUN_prof}, for the same B-field intensity, 
 reflecting the different CR source distribution.  
As expected, a steeper  CR distribution as in Fig. \ref{SUNLorim_prof} produces less synchrotron emission in the outer Galaxy, but more  emission in the inner Galaxy.
In the outer Galaxy, the difference in the synchrotron emission between the two distributions is a factor of 2. In particular, for the B-field model used, the agreement is better with the larger CR flux in the outer Galaxy; this would mean that there are more CR in the outer Galaxy than expected based on the standard source distribution and CR propagation. This is also confirmed by gamma-ray data \citep{diffuse2}.
We then perform the fit, and the best-fitting values are reported in Table \ref{Table2}.
Also, our $\chi^2$ values of $SUNE$ and $SUNLorimE$ show that there is a preference for the   \cite{strong2010} over the  \cite{lorimer} CR distribution, mostly caused by the flatter CR distribution in the outer Galaxy.  
This comparison is intended to illustrate the effect, as previously done also in gamma rays \citep{diffuse2}.
The best source distribution model will  be investigated in future in a dedicated study in conjunction with gamma-ray data.

\subsection{Effects of reacceleration} 

In SOJ2011, we found that the reacceleration models usually used to fit CR secondary/primary do not fit well the synchrotron spectrum at lower frequencies. 
This was also recently confirmed by \cite{dragon}.
We did not exclude reacceleration models entirely, but our study posed a challenge to be investigated.
 The problem was that reacceleration models produce more secondary positrons and electrons at low energies which cause an overprediction of the synchrotron emission at low frequencies. 
At these energies solar modulation plays a big role, making uncertain the determination of interstellar secondaries, which are used for distinguishing between models of propagation. 
As an example, we report here the results using a reacceleration model as used in  \citep{diffuse2} to fit gamma-ray data. 
In $SUNLorimerv33E$, Model 1 with halo height  4 kpc and the \cite{lorimer} source distribution are used, so that $SUNLorimerv33E$ can be directly compared with $SUNLorimE$. 
CR spectral parameter values and the diffusion coefficient are the same as used in \cite{diffuse2}. 
The $\chi^2$ values of the two models  (see Table \ref{Table2} ) favor the spatial distribution given by the  diffusion model $SUNLorimE$. 
We omit showing the profiles since the differences from Fig. \ref{SUNLorim_prof} are not much evident. 
Reacceleration affects significantly the electron and hence synchrotron spectra, as already found in SOJ2011.
Hence we confirm the results in SOJ2011 using a similar reacceleration model.
We are not concerned  here in finding a reacceleration model with different propagation parameters that can fit the synchrotron spectra. 
This will be investigated in a following paper together with gamma-ray data.  


\section{Full-sky maps}

\begin{figure*}
\centering
\includegraphics[width=0.8\textwidth, angle=0] {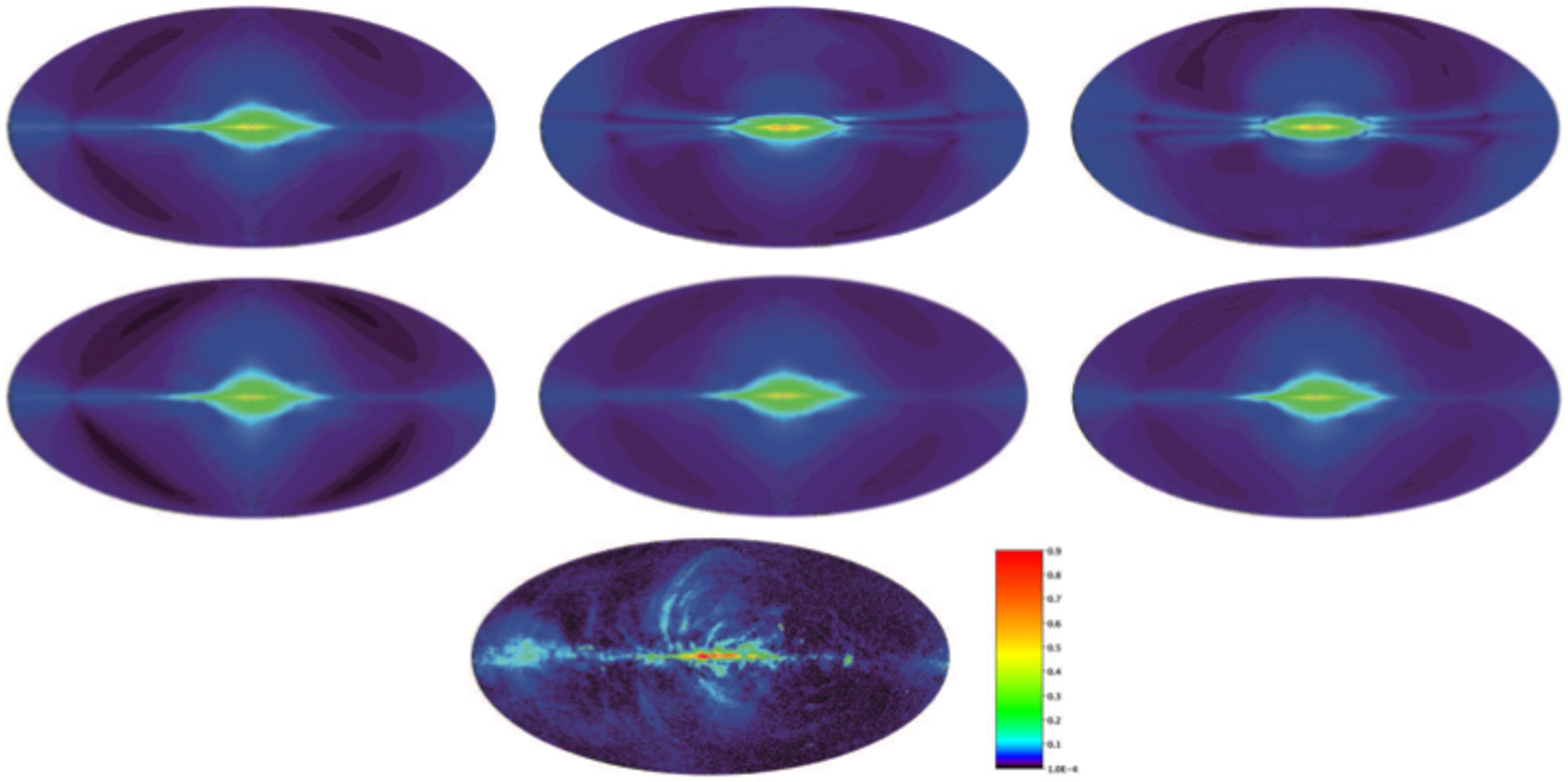}
\caption{$P$ 23 GHz sky maps. From top: SUNE, PSASSE,	PSBSSE,	SUN10E,	SUNLorimE and SUNLorimv33E. The sky map at the bottom shows {\it WMAP} data \citep{gold}. All sky maps have the same scale, with the Galactic longitude l = 0 in the centre. Units are mK.}
\label{P}
\end{figure*}

\begin{figure*}
\centering
\includegraphics[width=0.8\textwidth, angle=0] {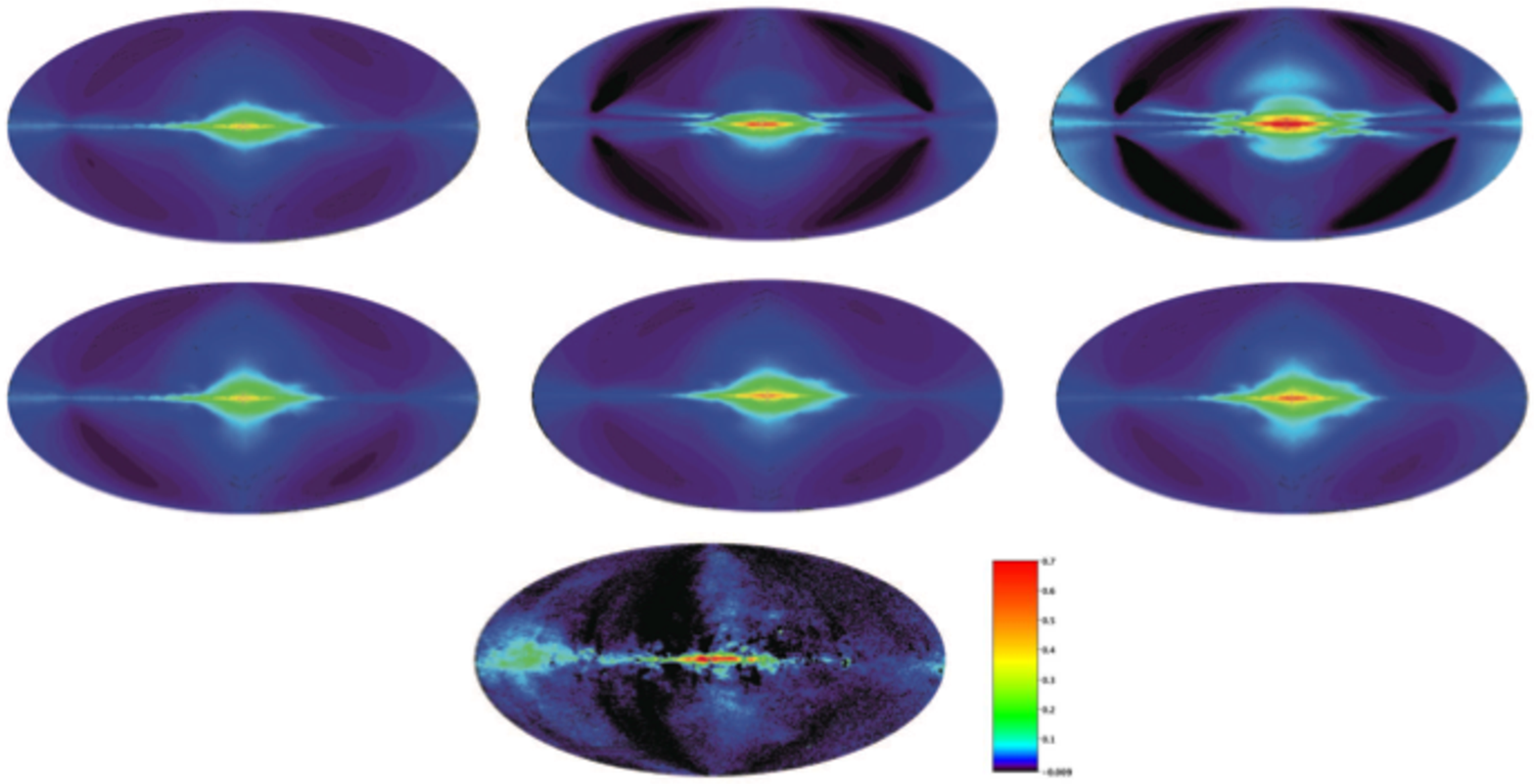}
\caption{$Q$ at 23 GHz sky maps. From top:  SUNE, PSASSE,	PSBSSE,	SUN10E,	SUNLorimE and SUNLorimv33E. The sky map at the bottom shows {\it WMAP} data \citep{gold}. All sky maps have the same scale, with the Galactic longitude l = 0 in the centre. Units are mK.}
\label{Q}
\end{figure*}

\begin{figure*}
\centering
\includegraphics[width=0.8\textwidth, angle=0] {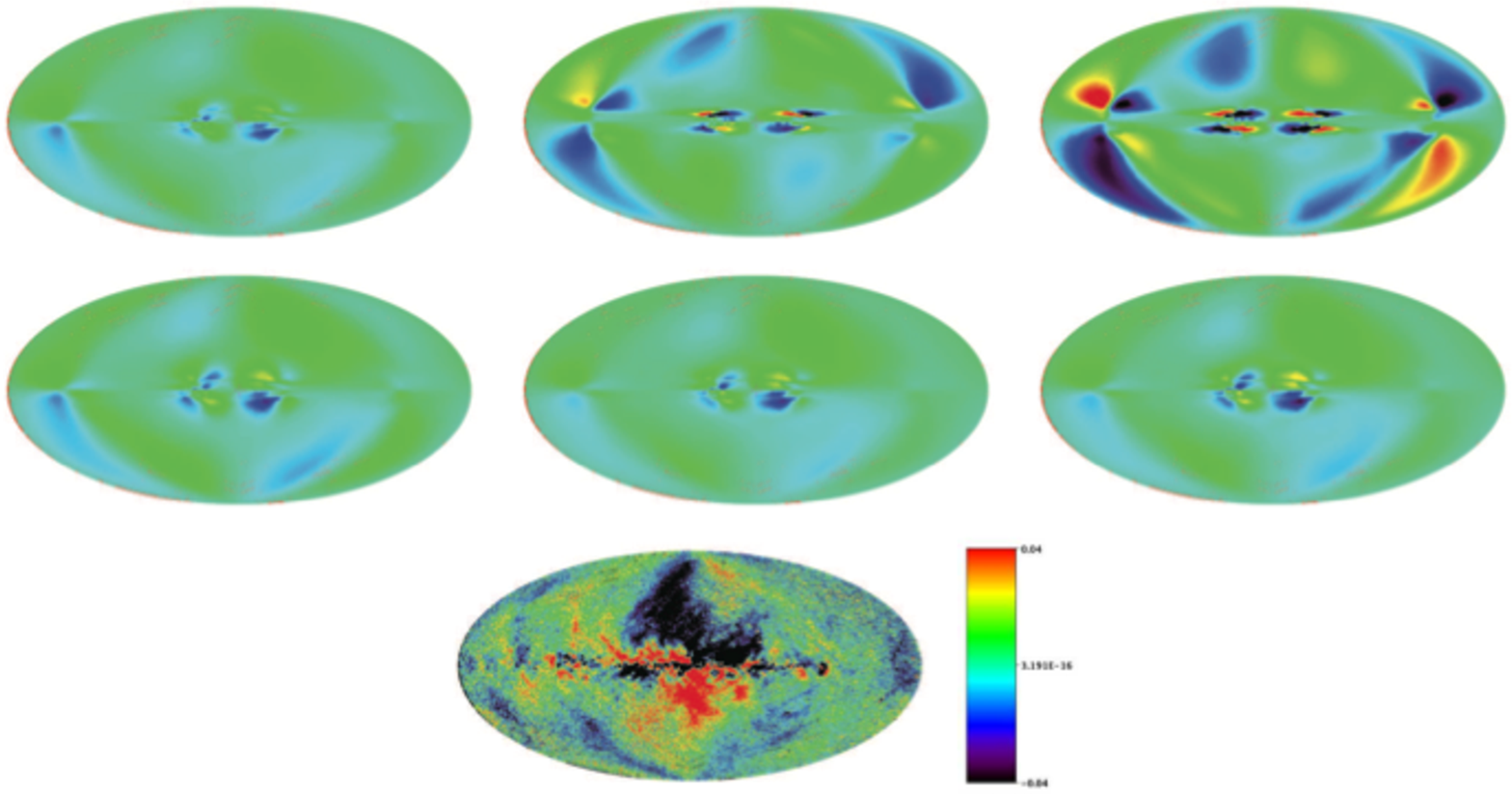}
\caption{$U$ at 23 GHz sky maps. From top: SUNE, PSASSE,	PSBSSE,	SUN10E,	SUNLorimE and SUNLorimv33E. The sky map at the bottom shows {\it WMAP} data \citep{gold}. All sky maps have the same scale, with the Galactic longitude l = 0 in the centre. Units are mK.}
\label{U}
\end{figure*}

Figures \ref{P}-\ref{U} show full sky maps\footnote{The sky maps were generated from HEALPix data sets using CDS ALADIN available from http:$\slash \slash$aladin.u-strasbg.fr; the GALPROP model synchrotron HEALPix maps were converted to a HEALPix format compatible with ALADIN using software available from http:$\slash \slash$www.mpe.mpg.de/$\sim$aws/propagate.html or https://sourceforge.net/projects/galpropskymapco.}
that give an overview of the global picture 
for 23~GHz $P, Q$ and $U$ for the all best-fitted models described in Table \ref{Table2} compared with the data. 
The $P$ sky maps in Fig \ref{P} show  a global correspondence between models and data, with the prominent exceptions of Loop I, the $l=145^o$ region and other specific features which are not included in these large-scale models. 
The $Q$ sky maps  in Fig \ref{Q} are mainly positive, with  negative regions around  $l= 90^o$ and $270^o$ 
corresponding to tangential viewing directions; 
the models do broadly resemble the {\it WMAP} data, indicating an agreement in global topology of the B-field. 
Considering  the $U$ sky maps in Fig~\ref{U}, the models do not bear much detailed resemblance to the data, and presumably they are dominated by structures not included in  such large-scale models.
In general, the models predict very characteristic positive/negative patterns in $U$ in the inner Galaxy and at high latitudes, which seem
 to be best matched by  $PSASSE$ and $PSBSSE$, and these patterns will deserve attention in future studies e.g. with the {\it Planck} satellite.

\begin{figure*}
\centering
\includegraphics[width=0.8\textwidth, angle=0] {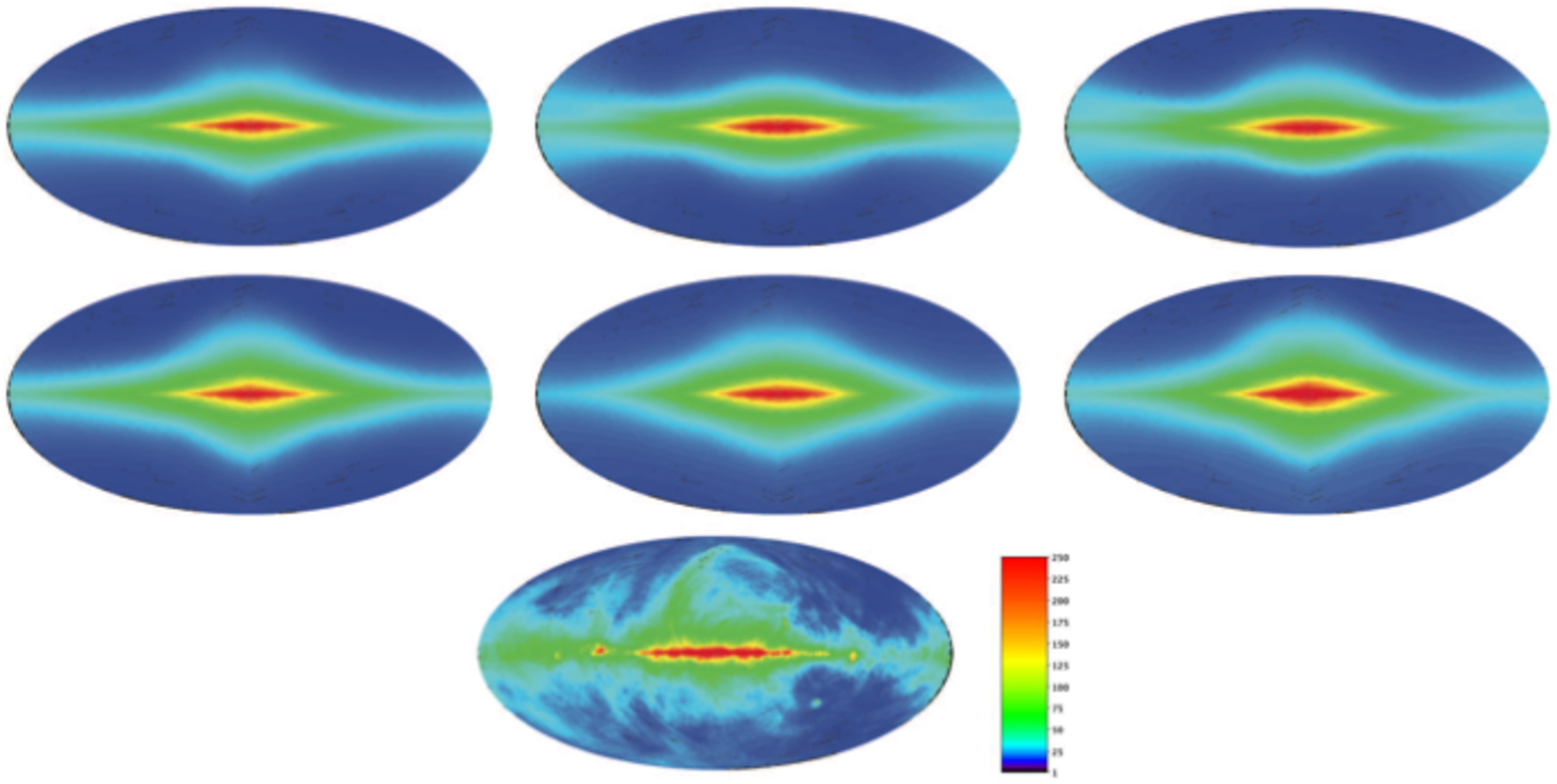}
\caption{408 MHz sky maps. From top: SUNE, PSASSE,	PSBSSE,	SUN10E,	SUNLorimE and SUNLorimv33E. The sky map at the bottom shows the data from \citet{haslam1982}. All sky maps have the same scale, with the Galactic longitude l = 0 in the centre. Units are K.  }
\label{408}
\end{figure*}

Figure \ref{408} shows the model and observed sky maps for 408~MHz. 
We see that all models reproduce roughly the large-scale structure of the emission, 
apart from Loop I, other features and localized sources, which are not included in our model. 
For the models used here, a flat CR source distribution such as in \cite{strong2010} resembles better the data for the outer Galaxy with respect to \cite{lorimer}. 
In all the sky maps shown here the offset as in Table~\ref{Table1} and \ref{Table2} are included.


\section{Potential shortcomings and improvements}
Throughout the paper we considered the anisotropic random field to be aligned with the regular field as in \cite{jansson2}. 
This  could result in a bad fit to  the polarization data.
CR propagation parameters could be different from the ones assumed here. 
In fact, the approximation that propagation parameters (such as the diffusion coefficient) of electrons are the same for electrons and protons could be inaccurate. 
Also, the assumption of isotropic homogeneous CR propagation  is probably  too simplistic.
The effects of  a Galactic wind \citep{Breitschwerdt} can be investigated as well.


\section{Conclusions}

We have described  extensions of the GALPROP code for synchrotron emission including polarization and absorption, 
and a preliminary model for free-free emission.
Using models for cosmic-ray propagation based on CR and gamma-ray data,
we investigated the consequences of changing different model parameters.
For the first time the synchrotron modelling has been treated in the context of CR propagation and realistic electron distributions and spectra, using polarized and total radio data to address the degeneracy among the different model parameters.  
Constraints coming from direct CR measurements and gamma-ray data are used. 
This work serves as baseline for future interpretations of radio data.

In summary:\\
1) We find that all-sky total intensity and polarization maps are reasonably reproduced by including an anisotropic random component of the B-field, which extends also in the halo.
 It has about the same intensity of the local regular component (1.5 - 2.3~\microG), given the regular B-field as constrained by RMs. \\
2) We study the sensitivity of the synchrotron modelling to different formulations of the regular B-field based on a small sample of B-fields from the literature.\\ 
3) We obtain a local random B-field  of 4.7 - 5.3 \microG. The scale lengths are 30 kpc in R and 4 kpc in z.\\ 
4) We choose the best-fitting model and use it as the sample for testing the sensitivity of our modelling to different propagation parameters.
 Increasing the halo height from 4 kpc to 10 kpc does not change significantly the emission in the plane, while a major effect is visible in the latitude profiles above $20^o$.  
 Changing the CR source distribution, for the same B-field intensity, rescales significantly the synchrotron model intensity as well.
We showed that in the outer Galaxy, the difference in the synchrotron emission between two distributions is a factor of 2, while in the inner Galaxy region is $\approx$ 20 per cent, reflecting the different CR source distribution. 
For the different models, a flat CR source distribution in the outer Galactic plan resembles the data best. 
This would support recent independent gamma-ray results.
The reacceleration models used in \cite{diffuse2} overestimate the synchrotron emission at low frequencies, confirming the finding in SOJ2011, that  present reacceleration models need to be critically evaluated.\\ 
5) Our analysis confirms the preference of a halo height larger than 4 kpc in order to fit the high-latitude data.\\

We conclude that propagation models and B-field models should be studied simultaneously, because both have influence  on the synchrotron modelling.  
In future, a parallel study of  gamma-ray and radio emission, together with CR measurements, can put better constraints on all the components involved.
In general, the improved knowledge on the Galactic B-field in a propagation modelling context and the effect of different CR electron propagation parameters, as provided in this work, represent a significance advance in the understanding of these subjects.
Our approach will be useful for the interpretation of Galactic emission observed by the {\it Planck} mission.
 It helps in component separation and for multiwavelength studies including gamma-rays observed by {\it Fermi}-LAT and  {\it INTEGRAL}. 
These studies are also relevant  for radio surveys which will be provided by  the Low-frequency Array and the forthcoming Square Kilometre Array telescope.



\clearpage

\section*{Acknowledgments}
We thank  Seth Digel, Igor Moskalenko, 
Troy Porter and Andrey Vladimirov for useful comments. Particular thanks are due to Gudlaugur J\'{o}hannesson.
We also thank Wolfgang Reich, Tess Jaffe and Anthony Banday for helpful discussions on an early draft of this paper.
We also acknowledge the anonymous referee for useful suggestions, which improved the paper. 

This work makes  use of HEALPix\footnote{http://healpix.jpl.nasa.gov/} described in \cite{healpix}.
E.O. acknowledges
support via NASA Grant No. NNX13AH72G and
NNX09AC15G.


\appendix

\section[]{Data used}

In this appendix, we give more details of the data and its usage.

\subsection{{\it WMAP} products}

The {\it WMAP}  files used from the LAMBDA site are as follows.

Polarized sky maps in {\it WMAP} frequency bands. These are smoothed (to $1^o$ resolution) order-9 HEALPix maps of $I, Q$ and $U$ with 3145728 equal solid angle sky pixels about $0.11^o$ on a side.
\begin{verbatim}

wmap_band_smth_iqumap_r9_7yr_K_v4.fits
wmap_band_smth_iqumap_r9_7yr_Ka_v4.fits
wmap_band_smth_iqumap_r9_7yr_Q_v4.fits
wmap_band_smth_iqumap_r9_7yr_V_v4.fits
wmap_band_smth_iqumap_r9_7yr_W_v4.fits

\end{verbatim}

Dust plus spinning dust templates:  
\begin{verbatim}

wmap_mcmc_sd_w_dust_temp_7yr_v4.fits       
wmap_mcmc_sd_w_dust_stk_q_7yr_v4.fits        
wmap_mcmc_sd_w_dust_stk_u_7yr_v4.fits        
wmap_mcmc_sd_k_spin_dust_temp_7yr_v4.fits

\end{verbatim}

\subsection{Noise bias in $P$}
Noise bias in $P=\sqrt(Q^2+U^2)$ (`P-bias') induced by noise in Q and U need to be addressed.
From the LAMBDA website and \cite{Jarosik}, the K-band $Q$, $U$ has pixel noise $\sigma_o=1.456$ mK;
since $\sigma=\sigma_o/\sqrt N_{obs}$, where  $N_{obs}\approx1000$ is the number of observations,
the bias in $P$ is therefore about  $1.456/\sqrt1000\times \sqrt2 = 0.065$ mK assuming $Q$ and $U$ are uncorrelated.
Since there is some $Q$, $U$ correlation, this value is just a rough estimate, but using the full correlation matrix is beyond what is required here.
To average $P$ over sky areas as required for spectra or profiles, we use  $\bar{P}=\sqrt(\bar{Q}^2+\bar{U}^2)$ where $\bar{Q},\bar{U}$ are the average of $Q$ and $U$ over pixels. 
This reduces the noise bias by a factor $\sqrt N_{pix}$, where  $N_{pix}$ is the number of pixels in the area;
since we average over typically $100^o$ in longitude for latitude profiles, so with the pixel size of $0.11^o$, we have
$N_{pix}\approx1000$, so the noise bias is reduced to a negligible level even compared to values near the Galactic poles, see the latitude profiles of P in Figs 5, 6, 7, 9 and 11. 
For longitude profiles along the Galactic plane $N_{pix}$ is smaller but the signal is much larger, so again the bias has little effect.
We are aware that averaging $Q$ and $U$ over large sky areas is not a rigorous procedure since the vector basis is changing with direction, but the data and models are treated in the same way so that at least the comparison is meaningful. In our fitting procedure, an offset is included when fitting $P$ in order to allow for the effect of P-bias.


.

\bsp

\label{lastpage}

\end{document}